\documentclass[twocolumn, twocolappendix]{aastex701}

\usepackage[svgnames]{xcolor}
\usepackage{multirow}
\definecolor{lapislazuli}{HTML}{3275A8}  % Tom mais comum

\hypersetup{
    colorlinks=true,
    linkcolor=lapislazuli,
    filecolor=magenta,
    anchorcolor=Salmon,      
    urlcolor=Crimson,
    citecolor=lapislazuli,
    }

\newcommand{\ocen}{$\omega$\,Cen}
\newcommand{\odwarf}{$\omega$\,Dwarf}

\usepackage{CJKutf8}
\usepackage[normalem]{ulem}
\usepackage[]{xcolor}
\usepackage{float}
\usepackage{amsmath}
\begin{document}
\begin{CJK}{UTF8}{gbsn}

\title{oMEGACat. X. Shedding light on the disrupted dwarf galaxy of Omega Centauri}

%%%%% Author list %%%%%%%%%%%%%%%%%%%%%%%%%%%%%%%%%%%%%%%%%%%%%%%%%%%%%%%%%%%%%%%%%%%%%%%%%%%%%%%%%%%%
\author[0000-0001-8052-969X]{S. O. Souza}
\affiliation{Max Planck Institute for Astronomy, K\"onigstuhl 17, D-69117 Heidelberg, Germany}
\email[show]{s-souza@mpia.de}  
\correspondingauthor{Stefano O. Souza}

\author[0000-0002-6922-2598]{N. Neumayer}
\affiliation{Max Planck Institute for Astronomy, K\"onigstuhl 17, D-69117 Heidelberg, Germany}
\email[]{neumayer@mpia.de}

\author[0000-0003-0248-5470]{A. C. Seth}
\affiliation{Department of Physics and Astronomy, University of Utah, Salt Lake City, UT 84112, USA}
\email[]{aseth@astro.utah.edu}

\author[0000-0003-2512-6892]{Z. Wang (王梓先)}
\affiliation{Department of Physics and Astronomy, University of Utah, Salt Lake City, UT 84112, USA}
\email[]{wang.zixian.astro@gmail.com}

\author[0009-0005-8057-0031]{C. Clontz}
\affiliation{Department of Physics and Astronomy, University of Utah, Salt Lake City, UT 84112, USA}
\affiliation{Max Planck Institute for Astronomy, K\"onigstuhl 17, D-69117 Heidelberg, Germany}
\email[]{clontz@mpia.de}

\author[0000-0002-5844-4443]{M. H\"aberle}
\affiliation{European Southern Observatory, Karl-Schwarzschild-Straße 2, 85748 Garching, Germany}
\email[]{maximilian.haberle@eso.org}

\author[0000-0002-2941-4480]{M. S. Nitschai}
\affiliation{Max Planck Institute for Astronomy, K\"onigstuhl 17, D-69117 Heidelberg, Germany}
\email[]{nitschai@mpia.de}

\author[0000-0002-7489-5244]{P. J. Smith}
\affiliation{Max Planck Institute for Astronomy, K\"onigstuhl 17, D-69117 Heidelberg, Germany}
\affiliation{Fakult\"at f\"ur Physik und Astronomie, Universit\"at Heidelberg, Im Neuenheimer Feld 226, D-69120 Heidelberg, Germany}
\email[]{pesmith@mpia.de}

\author[0000-0002-8077-4617]{T. Matsuno}
\affiliation{Astronomisches Rechen-Institut, Zentrum f\"ur Astronomie der Universit\"at Heidelberg, M\"onchhofstrasse 12–14, 69120, Heidelberg, Germany}
\email[]{matsuno@uni-heidelberg.de}

\author[0000-0002-1317-2798]{G. Guiglion}
\affiliation{Zentrum f\"ur Astronomie der Universit\"at Heidelberg, Landessternwarte, K\"onigstuhl 12, D-69117 Heidelberg, Germany }
\affiliation{Max Planck Institute for Astronomy, K\"onigstuhl 17, D-69117 Heidelberg, Germany}
\affiliation{Leibniz-Institut f{\"u}r Astrophysik Potsdam (AIP), An der Sternwarte 16, 14482 Potsdam, Germany}
\email[]{guiglion@mpia.de}

\author[0000-0002-0160-7221]{A. Feldmeier-Krause}
\affiliation{Department of Astrophysics, University of Vienna, T\"urkenschanzstrasse 17, 1180 Wien, Austria}
\email[]{anja.krause@univie.ac.at}

\author[0000-0002-6072-6669]{N. Kacharov}
\affiliation{Leibniz-Institut f{\"u}r Astrophysik Potsdam (AIP), An der Sternwarte 16, 14482 Potsdam, Germany}
\email[]{kacharov@aip.de}

\author[0000-0003-4546-7731]{G. van de Ven}
\affiliation{Department of Astrophysics, University of Vienna, T\"urkenschanzstrasse 17, 1180 Wien, Austria}
\email[]{glenn.vandeven@univie.ac.at}

\author[0000-0002-3651-5482]{J. Li}
\affiliation{Max Planck Institute for Astronomy, K\"onigstuhl 17, D-69117 Heidelberg, Germany}
\email[]{jdli@mpia.de}

\author[0000-0001-9673-7397]{M. Libralato}
\affiliation{Istituto Nazionale di Astrofisica - Osservatorio Astronomico di Padova, Vicolo dell’Osservatorio 5, Padova, IT-35122}
\email[]{mattia.libralato@inaf.it}

\author[0000-0003-3858-637X]{A. Bellini}
\affiliation{Space Telescope Science Institute, 3700 San Martin Dr., Baltimore, MD, 21218, USA}
\email[]{bellini@stsci.edu}

\author[0000-0001-7506-930X]{A. P. Milone}
\affiliation{Dipartimento di Fisica e Astronomia “Galileo Galilei”, Universita’ di Padova, Vicolo dell’Osservatorio 3, Padova, IT-35122}
\affiliation{Istituto Nazionale di Astrofisica - Osservatorio Astronomico di Padova, Vicolo dell’Osservatorio 5, Padova, IT-35122}
\email{antonino.milone@unipd.it}

\author[0000-0002-1212-2844]{M. Alfaro-Cuello}
\affiliation{Centro de Investigación en Ciencias del Espacio y Física Teórica, Universidad Central de Chile, La Serena 1710164, Chile}
\email[]{mayte.alfaro@ucentral.cl}

\begin{abstract}
Omega Centauri (\ocen{}) is the most massive and chemically complex star cluster in the Milky Way and is widely regarded as the surviving nuclear star cluster of an accreted dwarf galaxy. However, its parent host remains uncertain. Here, we investigate a scenario in which Sequoia, Thamnos, and Gaia--Enceladus (GE) are debris from a single disrupted progenitor, the \textit{\odwarf{}}, whose nucleus survives today as \ocen{}. Using APOGEE and GALAH abundances together with \textit{Gaia} astrometry, we reconstruct the chemical structure across this progenitor adopting orbital energy as a proxy for pre-merger radius. We find that the chemically evolved (younger Al-N-He-rich) population is strongly concentrated toward the inner regions, representing a population formed after/during the merger, while the primordial population represents a dwarf-galaxy-like population, supporting a common dwarf-galaxy origin for its components. The metallicity profile shows an inverted U-shaped gradient similar to those observed in present-day nucleated dwarfs. At the same time, the inner regions (\ocen{}+Thamnos) are more $\alpha$-enhanced than the outskirts, pointing to shorter and more efficient star formation and indicating that the nucleus may have assembled through the merger of inspiraling globular clusters. Neutron-capture abundances reveal a Eu-rich, r-process-dominated outskirts and inner regions enhanced in [Ba/Eu] and [La/Eu], requiring delayed enrichment and more complex chemical evolution. Finally, our analysis shows that Sequoia and Thamnos naturally fit an outside-in stripping sequence around \ocen{}, whereas the connection with GE remains unsure.
\end{abstract}

\keywords{\uat{Galaxies}{573} --- \uat{Nucleated dwarf galaxies
}{1130} --- \uat{Chemical abundances}{224} --- \uat{Star clusters}{1567} --- \uat{Milky Way dynamics}{1051} --- \uat{Galaxy mergers}{608}}

\section{Introduction}

The Milky Way (MW) assembled through a sequence of mergers between the proto-Galaxy and both massive and low-mass systems \citep[e.g.][]{Bullock2005,Kruijssen2019EMOSAIC,Fattahi2020,Aghanim2020}. {Strong evidence suggests that the earliest major merger occurred around $1.5\,Gyr$ after the MW formation \citep[Low-Energy/Kraken/Heracles, LKH, progenitor;][]{Massari2019,Kruijssen2020,Horta2021,Horta2023,GarciaBethencourt2023,Massari2026}. Additional mergers have continued to shape the Galactic structure over cosmic time \citep[e.g.][]{Ibata1994,Forbes2020,Helmi2020,Horta2023,Limberg2024,Deason2024}. However, because many of these events occurred at early epochs, their direct dynamical signatures in phase space have been largely erased \citep[e.g.][]{Pagnini2023}. } 

The Gaia-Enceladus system \citep[GE;][]{Helmi2018,Belokurov2018} is widely interpreted as the remnant of a massive dwarf galaxy that merged with the MW approximately $8–11\,Gyr$ ago \citep[e.g.][]{Villalobos2008, Montalban2021}. Astrometric measurements by the \textit{Gaia} satellite enabled the placement of stars with chemical abundances measured by APOGEE in the full six-dimensional phase space. Using this information, \cite{Helmi2018} showed that the inner stellar halo contains a slightly retrograde and chemically distinct population, most plausibly originating from a major merger that occurred $\sim10\,Gyr$ ago and contributed to the formation of the thick disc. Independently, \cite{Belokurov2018} reached similar conclusions, identifying a distinct, moderately metal-poor population ($-2.0<$[Fe/H]$<-1.0$) characterised by an elongated distribution in velocity space ($-100<v_\phi<+100\,\mathrm{km\,s^{-1}}$ and $-400<v_r<+400\,\mathrm{km\,s^{-1}}$), consistent with the accretion of a dwarf galaxy of mass $\sim10^{10}\,M_\odot$. From a chemical perspective, GE stars display a pronounced $\alpha$-enhanced plateau at low metallicity\footnote{We are assuming along this work the iron abundance [Fe/H] as the metallicity.}, along with distinctive Fe-peak and neutron-capture element patterns when compared to both the in-situ halo and classical dwarf satellites \citep[e.g.][]{Feuillet2021, Carrillo2022,Limberg2022, Fernandes2023}. %One-zone chemical-evolution models specifically tailored to GE \citep{Vincenzo2019,Hasselquist2021,Ernandes2024} consistently indicate low star-formation efficiency, rapid early enrichment, and a star-formation duration of $\gtrsim2$ Gyr, with quenching occurring shortly after infall.Dynamically, GE is characterised by relatively low binding energy, a small dispersion in angular momentum along the $z$-direction, and predominantly radial orbits \citep[][and references therein]{Limberg2022,Dodd2025}.

A second major retrograde accretion event is associated with the \textit{Sequoia} galaxy, identified as a high-energy, strongly retrograde component of the stellar halo in integrals-of-motion space \citep[e.g.][]{Koppelman2018,Myeong2018,Matsuno2019,Myeong2019}. {High-resolution abundance analyses of Sequoia candidates reveal that these stars define a systematically lower sequence in $\alpha$ and neutron capture elements than GE stars at $-1.8 \lesssim \mathrm{[Fe/H]} \lesssim -1.4$ \citep{Matsuno2022}. These trends point to an earlier onset of Type Ia supernovae (SNIa) and a slower overall enrichment history than in GE, consistent with a lower-mass progenitor.} Chemical-evolution interpretations further suggest that Sequoia experienced lower star-formation efficiency and higher outflow loading factors relative to GE, resulting in an earlier $\alpha$-knee and a lower mean metallicity \citep[e.g.][]{Monty2020,Matsuno2022,Fernandes2023,Ceccarelli2024}.

In addition to Sequoia, another substructure has been identified in the strongly retrograde halo, Thamnos. \citet{Koppelman2019} demonstrated that very retrograde halo stars can be separated into at least two dynamical components: a high-energy Sequoia group and a lower-energy, mildly retrograde population termed \textit{Thamnos}. Thamnos stars occupy lower-inclination, moderately eccentric retrograde orbits ($v_\phi \approx -150\,\mathrm{km\,s^{-1}}$) and are, on average, more metal-poor and more $\alpha$-enhanced than GE stars at similar metallicities. Early analyses suggested that the chemical properties of Thamnos resembled those of the in-situ thick disc, raising the possibility that Thamnos might represent dynamically heated disc stars rather than debris from an accreted dwarf galaxy \citep[e.g. ][]{Koppelman2019, Horta2023,Ceccarelli2025,Xie2026}. A colour-magnitude diagram analysis based on \textit{Gaia} photometry further indicates that Thamnos is, on average, older and more metal-poor, while Sequoia appears younger and more chemically evolved \citep{Dodd2025}.

{Studies based on simulations have proposed that Sequoia debris may represent the outer regions of the same merger event that produced GE \citep[e.g.][]{Koppelman2020,Amarante2022}. Thamnos stars, on the other hand, overlap chemically with both GE and Sequoia but retain distinctive signatures in $\alpha$, Fe-peak, and neutron-capture elements, implying related yet non-identical enrichment histories \citep{Koppelman2019, Horta2023, Ceccarelli2025}, indicating that Thamnos and Sequoia may constitute fragments of a larger retrograde system \citep[e.g.][]{Horta2023, Dodd2025, Ceccarelli2025}.}

Beyond their role in shaping the MW’s stellar halo, these accretion events have also been fundamental in assembling the present-day population of Galactic globular clusters \citep[GCs;][]{Forbes2010,Massari2019,Forbes2020,Kruijssen2019MERGERTREE,Kruijssen2020,Limberg2022,Souza2024,Massari2025}. Over the past decade, numerous studies have demonstrated that the majority of MW globular clusters (MWGCs) originated in external progenitor systems, with major mergers contributing more than half of the current GC population \citep{Forbes2010,Forbes2020,Myeong2018, Kruijssen2019MERGERTREE,Kruijssen2020}. 

More detailed investigations have sought to associate specific clusters--particularly those with high mass and broad metallicity distribution functions--as the likely nuclear star clusters \citep[NSCs;][]{Neumayer2020} of their progenitor dwarf galaxies \citep[e.g.][]{Pfeffer2021}. Within this framework, the clusters M19, M54, NGC~6934, and Omega Centauri (\ocen{}, NGC~5139) have been identified as the NSCs of the Kraken \citep{Kruijssen2020}, Sagittarius \citep{Ibata1994}, Helmi Streams \citep{Heimi1999}, and Gaia–Enceladus \citep[GE][]{Helmi2018,Belokurov2018} progenitors, respectively.

In particular, \ocen{} is the most massive and chemically complex GC in the MW and is widely interpreted as the remnant NSC of a disrupted dwarf galaxy \citep{Freeman1993,Tsuchiya2003,Limberg2022}. Its large stellar mass \citep[$3.55\times10^6\,M_\odot$;][]{Baumgardt2018}, the presence of a central intermediate-mass black hole \citep{Haeberle2024Nature}, retrograde orbital motion \citep{Limberg2022}, its broad metallicity range \citep[$-2.2 \lesssim \mathrm{[Fe/H]} \lesssim -0.6$;][]{Nitschai2023}, presence of tidal stream \citep[Fimbulthul;][]{Ibata2019}, and pronounced internal spreads in light, $\alpha$-, and neutron-capture elements clearly distinguish it from typical mono-metallic GCs \citep[e.g.][]{Johnson2010,Gratton2011,Marino2012,Clontz2024,Clontz2025}. 

Detailed abundance studies have revealed multiple chemically and kinematically distinct populations \citep[e.g.][]{Bellini2017,Clontz2026}, including metal-intermediate and metal-rich stars strongly enhanced in $s$-process elements, complex Na–O and Mg–Al anticorrelations, and unusual Fe-peak trends involving elements such as Mn and Cu \citep[e.g.][]{Johnson2009,Johnson2010, Marino2011, DOrazi2011, Romano2011,Garay2023}. Such properties are difficult to explain through self-enrichment in an isolated cluster but arise naturally if \ocen{} formed as the nucleus of a dwarf galaxy that underwent extended star formation, gas infall, and outflows prior to its accretion by the MW \citep[e.g.][]{Bekki2003,Romano2007, Marcolini2007}.

The identity of \ocen{}’s progenitor and its connection to the aforementioned retrograde structures remain subjects of active debate \citep{Myeong2018,Limberg2022,Pagnini2025,Dondoglio2025}. Based on \textit{Gaia} DR2 kinematics, \citet{Massari2019} associated the majority of inner-halo GCs with specific accretion events, assigning \ocen{} to GE on the basis of its integrals of motion and relatively high binding energy. Because NSCs are the most bound parts of galaxies, and are themselves massive objects, they experience more dynamical friction than other debris from a tidally disrupted system. This interpretation has been reinforced by dynamical modelling and black-hole scaling relations, which suggest that \ocen{}’s mass, age–metallicity relation, and putative intermediate-mass black hole are most naturally explained if it were once the nucleus of a GE-like dwarf galaxy with a mass comparable to that of the LMC \citep[e.g.][]{Pfeffer2021, Callingham2022, Limberg2022, Haeberle2024Nature, Limberg2024}. Conversely, \ocen{} falls naturally within the Sequoia region of the action-diamond plane, while Thamnos appears to represent more recent debris in integrals-of-motion space \citep{Myeong2018,Myeong2019, Koppelman2019}.

\begin{figure*}[ht!]
    \centering
    \includegraphics[width=1\linewidth]{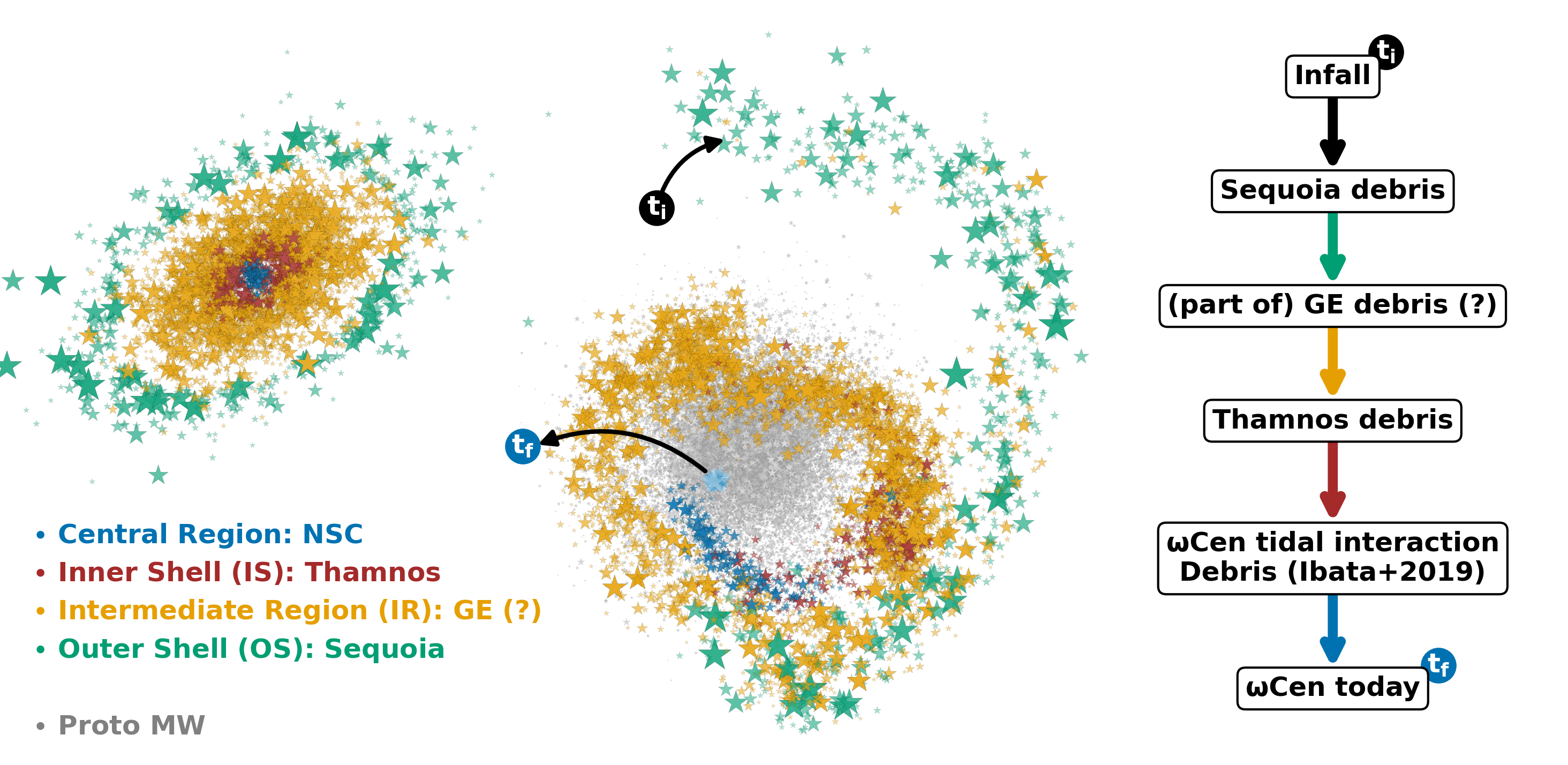}
    \caption{Schematic illustration of the proposed \odwarf{}. 
        On the left, the structure of the progenitor \odwarf{} galaxy is displayed, with the disc structure composed of
        an outer stellar component associated with Sequoia (green), an intermediate 
        component corresponding to Gaia--Enceladus (yellow), an inner component 
        identified with Thamnos (red), and a dense central nuclear star cluster (blue). 
        In the middle, an illustration of the disruption of this system after its infall 
        into the proto-Milky Way ($\rm t_i$), producing debris associated with each halo substructure mentioned before, as well as the tidal debris of 
        \ocen{} observed today ($\rm t_f$). The infall process is also explained in the right as a flowchart. Note: for illustration purposes, not to scale. This figure is inspired by figures in \cite{Koppelman2020} and \cite{Skuladottir2025}.} 
    \label{fig:scenario}
\end{figure*}

{In this work, we evaluate chemodynamical data on Sequoia, GE, Thamnos, and \ocen{} in the context of a unified model (\autoref{fig:scenario}). In this scenario, the inner part is composed of the past NSC, which survived as \ocen{}, and an inner shell linked to the nucleus through more recent tidal stripping represented by Thamnos. GE would correspond to an intermediate and enriched region that dominates the inner halo, while Sequoia would trace the weakly bound outer component populating the high-energy retrograde halo. We refer to this hypothetical progenitor as the \textit{Omega-Dwarf} (\odwarf{}). In Section~\ref{sec:selection} we compile and homogenise high-resolution abundance data for these retrograde substructures and for \ocen{} from the two most widely used spectroscopic surveys, APOGEE and GALAH \citep{Majewski2017, DeSilva2015}. In Section~\ref{sec:rise_odwarf} we detail supporting arguments for the one dwarf-galaxy hypothesis. In Section~\ref{sec:discussion}, we explore the present orbital energy as a proxy for the pre-merger radius to compare the abundance patterns in key diagnostic planes proposing the chemical enrichment history for \odwarf. The summary is drawn in Section~\ref{sec:summary}. }

\section{Target Selection} \label{sec:selection}

The central goal of this work is to characterise the debris field associated with the putative progenitor of \ocen{}. %, which we refer to as the \odwarf{}. 
To do so, we require a stellar sample with reliable chemical abundances and precise astrometry. %Our parent catalogues combine high-resolution spectroscopy from APOGEE DR17 and GALAH DR4 with astrometric measurements from \textit{Gaia} DR3. 
In the following, we describe the construction of this dataset and the methodology used to {assign membership probabilities to the three halo substructures -- Gaia–Enceladus, Sequoia, and Thamnos -- as well as to \ocen{} itself. Rather than applying binary hard cuts, we adopt a probabilistic approach based on Monte Carlo sampling that accounts for measurement uncertainties and yields a continuous membership probability for each star}.

\subsection{APOGEE DR17 parent sample} \label{sec:apogee}

In this work, we use the publicly available 17th data release (DR17) of APOGEE \citep{Majewski2017, Abdurrouf2022}. APOGEE DR17 provides near-infrared, high-resolution ($R\sim22{,}500$) H-band spectra ($1.5140$--$1.6940\,\mu$m) and elemental abundances for 20 species, spanning CNO, odd-$Z$, $\alpha$\footnote{Elements formed from fusion of Helium nuclei: Mg, Si, S, K, Ca}, iron-peak, and the $s$-process element Ce. Line-of-sight velocities are obtained from the reduction pipeline described by \citet{Nidever2015}, while atmospheric parameters and elemental abundances are derived with ASPCAP \citep{GarciaPerez2016}. In this study, we only consider the following high-quality abundances from ASPCAP: C, N, O, Mg, Si, Ca, Al, Mn, Fe. To ensure robust stellar parameters and abundances, we apply a set of standard quality-control cuts:

\begin{itemize}
\item Signal-to-noise ratio $S/N > 50$ pixel$^{-1}$,
\item No pipeline warning flags (\texttt{STARFLAG = 0} and \texttt{ASPCAPFLAG = 0}),
\item All element-specific quality flags for Fe (\texttt{FE\_H\_FLAG}), Mg, Mn, Al, N, Ni, and O (\texttt{X\_FE\_FLAG}) set to zero,
\item Restriction to giant stars (\texttt{TEFF\_SPEC}$ > 4200,{\rm K}$, \texttt{LOGG\_SPEC}$< 3.5$).
\end{itemize}

The APOGEE sample is cross-matched with \textit{Gaia} DR3 using \texttt{GAIA\_SOURCE\_ID}. To retain only high-quality astrometric solutions, we require \texttt{RUWE} $\le 1.4$ and \texttt{parallax\_over\_error} $>2$ \citep{Lindegren2021}. To homogenise the samples, we adopt distances from \texttt{StarHorse} \citep{Queiroz2018,Queiroz2020,Queiroz2023}, retaining only stars with fractional distance uncertainties $<20\%$. After the quality cuts, the number of stars in the APOGEE parent sample is $297,104$.

\subsection{GALAH DR4 parent sample} \label{sec:galah}

To complement APOGEE’s NIR abundances, we use the high-resolution GALAH DR4 \citep[$R\sim42,000$][]{Buder2025}, which spans a broad region of the optical wavelength range: $4713-4903$\,\AA{} (blue CCD or CCD1); $5648 – 5873$\,\AA{} (green/CCD2); $6478–6737$\,\AA{} (red/CCD3); and $7585–7887$\AA{} (IR/CCD4). This enables abundance measurements for all elements as those provided by APOGEE, and additionally, for example, Ba, La, and Eu, which we will consider in the following analysis.

Since the GALAH abundance pipeline was trained primarily on field stars \citep{Buder2025}, some abundance estimates--particularly for GC stars--can be unreliable. We identify a dense locus of stars with [Eu/Fe]~$\sim0.0$ that are not flagged but exhibit clear problems in the Eu abundance determination \citep{Kane2025}. We found that these stars with [Eu/Fe]~$\sim0.0$ also show higher [Cu/Fe] values than the overall population. Therefore, in addition to the recommended spectroscopic quality filters, we implement further cuts to remove unflagged problematic cases:

\begin{itemize}
\item \texttt{SP\_FLAG = 0},
\item $\texttt{SNR\_PX\_CCD3} > 30$,
\item $T_{\rm eff} > 4200,{\rm K}$ and $\log g > 1.0$,
\item \texttt{flag\_eu\_fe} set to 0 and \texttt{|eu\_fe|}$>0.001$,
\item \texttt{cu\_fe}$<0.0$.
\end{itemize}

With these additional cuts, we remove the majority of the unflagged Eu outliers, although some remain.

All GALAH stars are matched to \texttt{StarHorse} using their \texttt{gaia\_source\_id} and are subject to the same astrometric cuts adopted for APOGEE. The total number of stars in the final GALAH DR4 parent sample is $501,026$. %{From the final GALAH parent sample, $15,343$ stars have also APOGEE data.}

\subsection{Globular Cluster stars cleanning}
{To identify stars that are members of known globular clusters (other than \ocen), we use the \textit{Gaia} EDR3 catalogues of \cite{Vasiliev2021}, which provide membership probabilities based on proper motions, parallax, and sky position. We remove from both parent samples all stars with membership probability $>99\%$ in any known GC, except those associated with \ocen{} (which we treat separately, see \S\ref{sec:ocen_prob}). This cleaning ensures that our field sample is not contaminated by bound cluster members. The final APOGEE and GALAH parent samples contain $293,212$ and $500,357$ stars, respectively, after removing GC stars.}

\subsection{Orbital Parameters} \label{sec:orbit}

After constructing the cleaned parent samples, we derive orbital parameters to identify halo substructures and \ocen{} stars. For APOGEE and GALAH, we compute full six-dimensional phase-space coordinates and integrate the orbits forward for $10\,$Gyr in the MW potential of \citet{McMillan2017} using the Action-based GAlaxy Modelling Architecture \citep[\texttt{AGAMA};][]{Vasiliev2019} \texttt{Python} code. Although GALAH DR4 provides orbital parameters based on the \citet{McMillan2017} potential, those orbits adopt distances from \cite{BailerJones2021}. We therefore recompute the orbital parameters for GALAH DR4 to ensure consistency with our choice of distance dataset.

For each star, we compute the pericentre and apocentre distances, the specific orbital energy $E$, the azimuthal angular momentum $L_z$, and the orbital actions $(J_R, J_\phi, J_Z)$ following the conventions of \citet{Binney2008}. {Uncertainties are estimated from 100 Monte Carlo realisations, drawing from the error distributions of proper motions, radial velocities, and distances, and propagating them through the orbit integration. The median of the resulting distributions is adopted as the final value, and the spread is used to characterise uncertainties.} Following standard practice \citep[e.g.][]{Helmi2020}, we apply a broad halo pre-selection by removing stars with highly bound or unbound energies, or strongly prograde angular momentum.

\subsection{{Membership Probabilities}} \label{sec:subs}

{We now assign membership probabilities to the GE, Sequoia, and Thamnos, as well as to \ocen{} itself. For each substructure, we define a region in chemo-dynamical parameter space based on criteria established in the literature. Membership probabilities are computed via Monte Carlo sampling that propagates measurement uncertainties through the selection criteria.}

{For each star, we generate $N=1000$ realisations of its abundances, $E$, $L_Z$, $J_R$, $J_\phi$, and $J_Z$, drawn from the orbital integration. For each realisation, we evaluate whether the star satisfies the substructure-specific criteria defined below. The membership probability for a given substructure is the fraction of realisations that meet all criteria}:
\begin{align}
\displaystyle \mathcal{P}_{\rm substructure} = \frac{N_{\rm satify\, selection\, criteria}}{N}. \label{eq:prob_general}
\end{align}

\subsubsection{Gaia–Enceladus}

GE is the most extensively studied halo substructure. Following its discovery, many authors have proposed refined selections for GE debris. \cite{Carrillo2023} compiled commonly used criteria for GE selection and examined their implications by comparison with simulations of GE-like mergers. Frequently used selections in integrals-of-motion space (the $E$–$L_Z$ plane) adopt a box at low binding energy and near-zero angular momentum \citep[e.g.][]{Horta2023}. However, such cuts can retain a substantial fraction of in-situ stars. \cite{Carrillo2023} found that criteria based on radial action and angular momentum \citep[$J_r$–$L_Z$;][]{Feuillet2021,Limberg2022} yield a purer GE sample. {We therefore define the GE region as}:
\begin{align}
{\rm GE\ region} = \left\{ \begin{array}{rcccl}
-500 & \leq & L_z &\leq &+500\,{\rm kpc\,km\,s^{-1} } \\
30 &\leq &\sqrt{J_R} &\leq &50\,({\rm kpc\,km\,s^{-1} })^{1/2}
\end{array} \right. \label{eq:gse_def}
\end{align}

\subsubsection{Sequoia \& Thamnos}

In velocity space ($\sqrt{v_R^2 + v_z^2}$ vs. $v_\phi$), Thamnos and Sequoia correspond to the retrograde low-velocity and the so-called arch structures, respectively \citep[][]{Helmi2018}. Sequoia stars occupy high-energy, strongly retrograde, and vertically extended orbits, often characterised in the diamond action plane \citep{Myeong2019}. {We therefore define the Sequoia region as}:
\begin{align}
{\rm Sequoia\ region} = \left\{ \begin{array}{rl}
E & > -1.5\times10^5\ {\rm km^2\,s^{-2}}, \\
L_z/J_{\rm tot} & < 0, \\
J_Z/J_{\rm tot} & > 0.5, \\
\frac{J_Z - J_R}{J_{\rm tot}} & < 0.1
\end{array} \right. \label{eq:seq_sel}
\end{align}
where $J_{\rm tot}=J_R + |J_\phi| + J_Z$.

To select Thamnos stars, we adopt the criteria defined in the original work \citep{Koppelman2019} and used in \cite{Horta2023}:
\begin{align}
{\rm Thamnos\ region} = \left\{ \begin{array}{rl}
L_z & < 0, \\
-1.8 & < E < -1.6 \ (10^5\ {\rm km^2\,s^{-2}}), \\
ecc&<0.7, \\
&\text{not in other region.}
\end{array} \right. \label{eq:tham_sel}
\end{align}
where $ecc$ is the orbital eccentricity defined as $(r_{\rm apo}-r_{\rm peri})/(r_{\rm apo}+r_{\rm peri})$.

\subsubsection{{Omega Centauri}} \label{sec:ocen_prob}

{To select \ocen{} members, we directly adopt the membership probabilities from the \textit{Gaia} EDR3 catalogue of \cite{Vasiliev2021}. %, which provides probabilities based on proper motions, parallax, and sky position for stars in each known globular cluster. 
For each star in our parent samples that is matched to this catalogue, we set $\mathcal{P}_{\rm \omega Cen} = P_{\rm Vasiliev}$. Stars not present in the Vasiliev catalogue are assigned $\mathcal{P}_{\rm \omega Cen} = 0$, and assumed those with $P_{\rm Vasiliev}>0.999$ (about 62,000 stars from the original catalogue) as the golden sample and true members of \ocen{}. The final number of \ocen{} members is in \autoref{tab:pop_numbers}.}

\subsection{{Final \odwarf{} Membership Probability}} \label{sec:final_prob}

{For each star, we compute $\mathcal{P}_{\rm GE}$, $\mathcal{P}_{\rm Sequoia}$, and $\mathcal{P}_{\rm Thamnos}$ using Equation~\ref{eq:prob_general} with the respective region definitions and $\mathcal{P}_{\rm \omega\, Cen}$ from \cite{Vasiliev2021} catalogue. We then define the probability of belonging to the \odwarf{} as the combined probability of membership in any of the four substructures, accounting for mutual exclusivity}:

\begin{align}
\mathcal{P}_{\rm \omega\,Dwarf} &=
\mathcal{P}_{\rm GE} + \mathcal{P}_{\rm Sequoia} + \mathcal{P}_{\rm Thamnos} + \mathcal{P}_{\rm \omega\,Cen} \nonumber \\
&\quad
- \sum_{i\,=!\,j} \mathcal{P}_{i \cap j}
\label{eq:final_prob}
\end{align}
{where the intersection probabilities $\mathcal{P}_{i \cap j}$ account for the fact that a star cannot simultaneously belong to two distinct substructures.}

\begin{figure*}[ht!]
    \centering
    \includegraphics[width=0.99\linewidth]{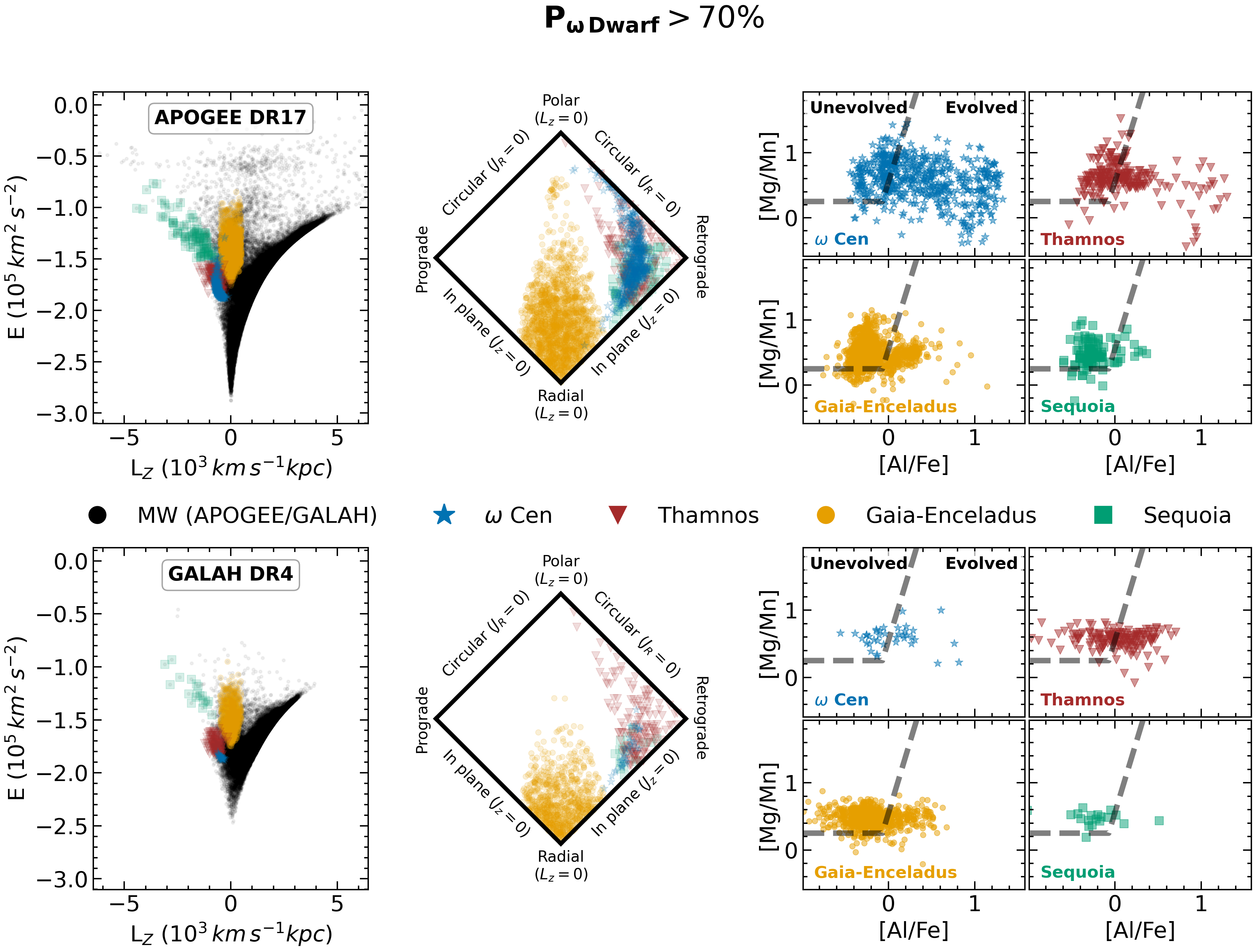}
    \caption{ \odwarf{} stars  with $P_{\omega\,\mathrm{Dwarf}} > 70\%$. The top row shows the APOGEE sample, and the bottom row shows the GALAH sample. The left panels are the orbital energy ($E$) as a function of the vertical component of angular momentum ($L_z$). Black points trace the MW population. Colored symbols identify stars in Sequoia (green squares), Gaia--Enceladus (orange circles), Thamnos (red triangles), and $\omega$ Cen (blue stars). The middle panels present the orbital-action space in a diamond projection. The right panels show the chemical plane [$\mathrm{Mg}/\mathrm{Mn}$] versus [$\mathrm{Al}/\mathrm{Fe}$], with unevolved stars in the left subpanels delimited by the dashed line and evolved stars in the right subpanels. }
    \label{fig:target_selection}
\end{figure*}

{We consider stars with $\mathcal{P}_{\rm \omega\,Dwarf} > 0.7$ likely members of the \odwarf{}, though we retain the full probability distribution for subsequent analyses to properly account for selection uncertainties. The number of stars with $\mathcal{P}_{\rm \omega\,Dwarf} > 0.7$ in each parent sample likely belong to each substructure is listed in \autoref{tab:pop_numbers}, and the selection is illustrated in \autoref{fig:target_selection}, which shows the median positions of high-probability members in the relevant chemo-dynamical planes.}

\begin{deluxetable}{lccc}[h!]
\tablecaption{{Number of stars in the parent samples and in \odwarf{} components.}\label{tab:pop_numbers}}
\tablehead{
\colhead{} & 
\colhead{APOGEE} &
\colhead{GALAH} &
\colhead{A$\cap$G}
}
\startdata
Parent Sample  & 249,699 & 500,357 & 15,343 \\
\hline
\ocen{}        & 697  & 179  & 58  \\
Thamnos        & 263  & 531  & 20  \\
Gaia-Enceladus & 1973 & 1570 & 107 \\
Sequoia        & 123  & 85   & 7  \\
\hline
\odwarf{} & 3056 & 2365 & 192 \\
\enddata
\tablecomments{\footnotesize (a) sample; (b) numbers related to APOGEE; (c) GALAH; (d) intersection between APOGEE and GALAH. The stars in each \odwarf{} components have membership probability $\mathcal{P_{\rm \omega\,Dwarf}}>0.7$ and associate to the component for which the individual probability is maximum. }
\end{deluxetable}

\subsection{Abundance consistency}

\begin{figure*}[hbt!]
    \centering
    \includegraphics[width=0.99\linewidth]{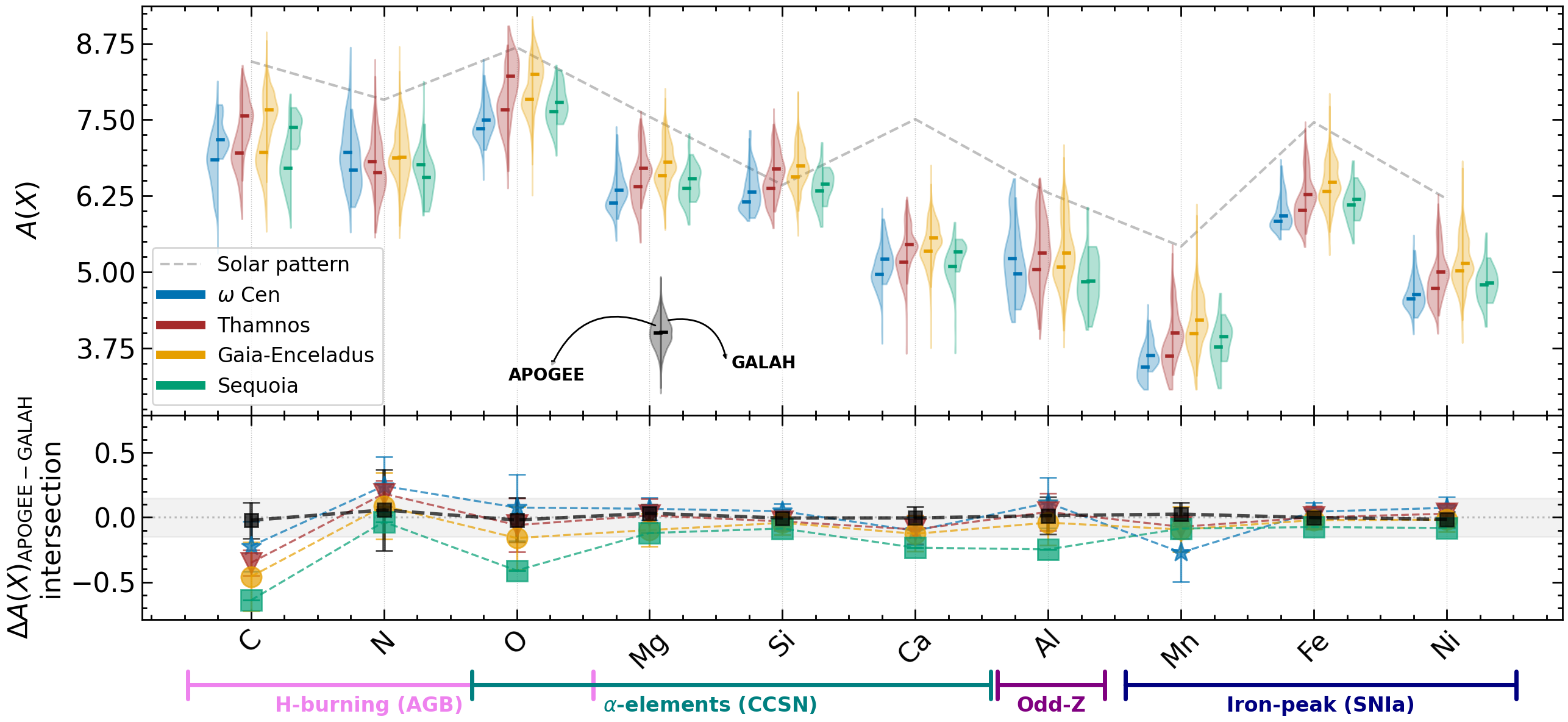}
    \caption{Chemical pattern of \odwarf{} for the elements provided by both APOGEE and GALAH. Upper panel: the violins represent the distribution obtained from the data. For each element, the left panel shows the distribution from APOGEE, while the right panel shows that from GALAH. The distribution for each \odwarf{} substructure is represented using the same colour-code as in \autoref{fig:target_selection}. The dashed grey line is the solar chemical pattern. Bottom panel: the star-by-star median difference for the intersection sample APOGEE--GALAH. The black squares represent the medians considering all intersection stars, while the colored symbols and lines show the medians for the individual populations.}
    \label{fig:violin}
\end{figure*}

To ensure consistency between the two surveys for our sample, we compare the abundance distributions of the elements in common between both surveys. The bottom panel of \autoref{fig:violin} shows the median difference star-by-star for the intersect sample (last column of \autoref{tab:pop_numbers}). The global median difference distribution, considering all stars, is in black square, while the distributions considering only the stars in each population are in blue-star (\ocen{}), red-triangle (Thamnos), yellow-dot (GE), and green-square (Sequoia). The global comparison shows median values well confined in 0.10 dex, representing no substantial difference between surveys. This result is similar to that of \cite{Hegedus2023}, who performed a comparative analysis of APOGEE DR17, GALAH DR3, and the Gaia-ESO survey. Their APOGEE--GALAH sample is composed of $15,537$, a similar number found in this work. Conversely, within the individual populations, the differences exceed 0.10 dex for C, N, O, Al, and Mn. These elements trace the chemical evolution of the stellar population, indicating contamination by evolved stars. Since the GALAH DR4 data-driven algorithm is trained on MW field stars, the evolved population of the Galactic disc may explain the median differences. 

The upper panel of \autoref{fig:violin} shows the abundance distribution for each element separated by \odwarf{} components. For each element, the left half shows APOGEE data, and the right half shows GALAH data. As shown in the bottom panel, for the reduced-intersect sample, the median differences for Mg, Si, Ca, Fe, and Ni are all less than 0.10 dex. Comparing the full two samples in violin plots (upper panel), these elements show that the GALAH distributions seem to be confined within the APOGEE ones. Even though all elements show a systematic offset between APOGEE and GALAH, the pattern observed in APOGEE for the individual populations is also present in GALAH. For example, apart from the mean value, Fe is lower for \ocen{}, increases for Thamnos and GE, and decreases again for Sequoia. This consistency in chemical patterns ensures that the global results we find in the following sections are independent of the survey. For those reasons, for the elements listed in \autoref{fig:violin}, we will focus on APOGEE data and employ GALAH for the neutron-capture elements Ba, La, and Eu. The only exception will be the metallicity [Fe/H], for which we will, for completeness, analyse with both surveys.

\section{One dwarf galaxy hypothesis: \\The rise of \odwarf{}}\label{sec:rise_odwarf}

The one dwarf galaxy hypothesis argues that \ocen{}, Thamnos, GE, and Sequoia are not independent accretion events, but rather correspond to different layers of the same disrupted system, deposited at different stages of its orbital decay. The key mechanism governing this evolution is the coupled action of tidal stripping and dynamical friction which extracts orbital energy and angular momentum from the infalling dwarf galaxy \citep{Chandrasekhar1943,Binney2008}. In practice, however, the efficiency of dynamical friction is strongly time-dependent because the satellite is being stripped at the same time it keeps orbiting the MW. As the outer parts are removed outside-in, the bound mass of the remnant drops, lengthening the dynamical-friction timescale and making the subsequent decay increasingly inefficient. This coupling between tidal stripping and friction provides an energy-radius ordering from which the first material to be unbound comes from large radii in the progenitor and is deposited on the lowest-binding-energy (highest E) orbits, whereas later passages peel off progressively more tightly bound layers that end up at higher binding energies (lower E) in the MW potential. Indeed, such a scenario has already been investigated for GE by \citet{Skuladottir2025} and \citet{Berni2026}, who found evidence of multiple passages in the GE debris. 

\cite{Koppelman2020}, using simulations by \cite{Villalobos2008, Villalobos2009}, attempted to link the halo phase-space properties to a major merger involving a GE-like dwarf galaxy. They compared a suite of 12 simulations--six prograde and six retrograde encounters--equally split between spheroidal and discy progenitors. They found that reproducing both the arch and the retrograde structures in halo phase space favours a merger with a retrograde, disc-like dwarf on an orbit inclined by $\sim30^\circ$ with respect to the Galactic plane. They also showed that the first stars captured during the interaction originated in the dwarf’s outskirts and today populate the most retrograde, lowest-binding-energy region of the $E$–$L_Z$ plane. As disruption proceeds, progressively more tightly bound material is stripped, such that the remnant debris shifts toward nearly zero $L_Z$ and higher binding energy, corresponding to stars deposited deeper in the Galactic potential.

On the other hand, \cite{JeanBaptiste2017} and \citet{Pagnini2023}, using $N$-body simulations of 1:10 major mergers, showed that, several Gyr after a retrograde encounter, the debris can occupy a broad region of the $E$–$L_Z$ plane. At the same time, persistent overdensities may still be visible even $\sim$5 Gyr after the merger. This result aligns with the AURIGA simulation employed in \citet{Skuladottir2025}. In their Figure 1, the density distribution of stars accreted from GE spans the entire $E$–$L_Z$ space, showing overdensities at the mean loci of GE and Sequoia. 

A remaining related question would be why the NSC of this dwarf galaxy -- identified with \ocen{} -- has not inevitably spiralled into the Galactic center. The inspiral can effectively stall once the satellite has been reduced to a compact, less massive nucleus if severe tidal stripping can remove most of the dwarf mass before dynamical friction has time to drive the remnant to the MW center, leaving a naked nucleus on a non-zero pericentre orbit and a longer friction timescale \citep{Bekki2003, Tsuchiya2003,Read2006,Petts2015,Petts2016}. 

\begin{figure}[htb!]
    \centering
    \includegraphics[width=0.99\linewidth]{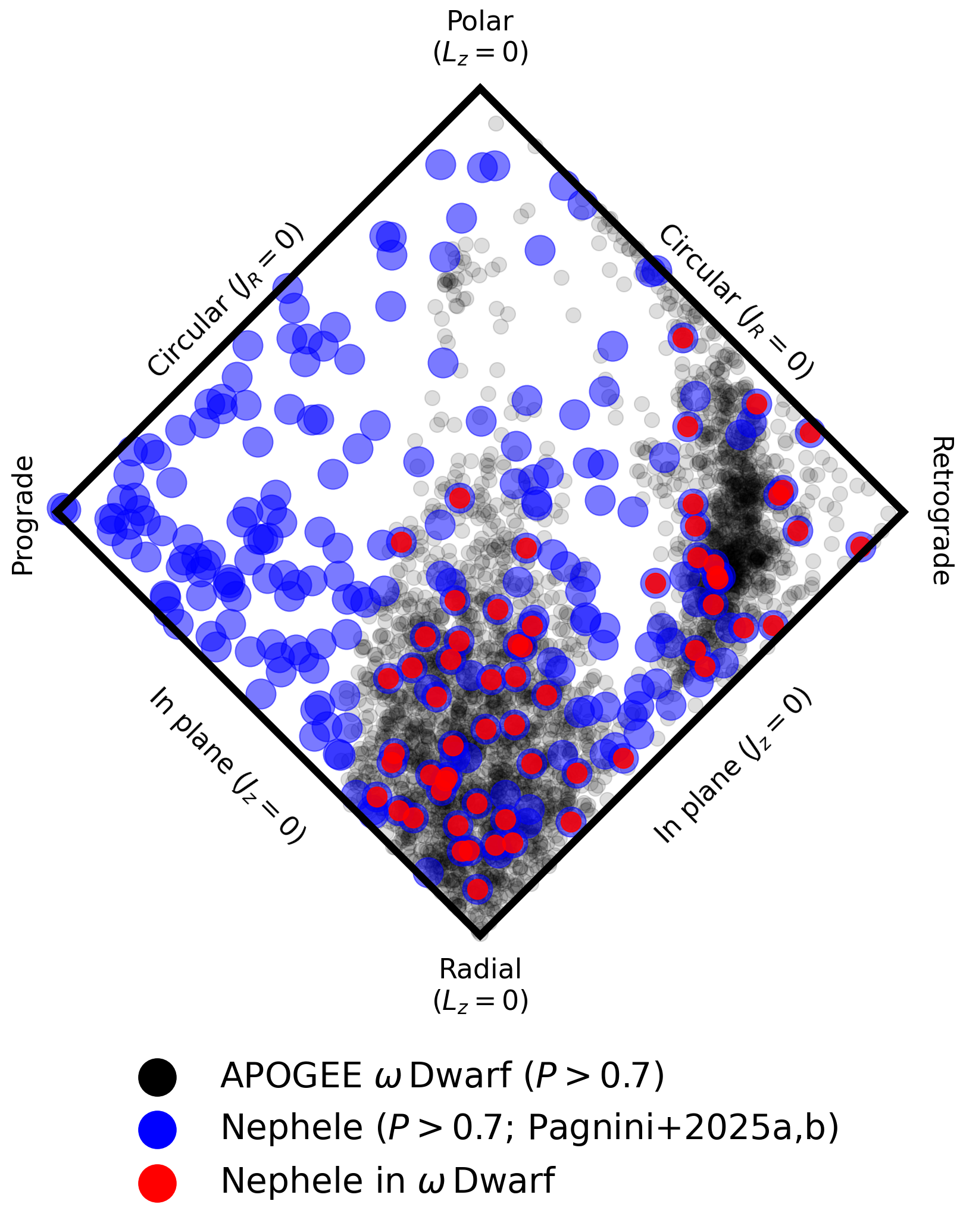}
    \caption{Orbital distribution in action space for stars associated with \odwarf{} and Nephele \citep{Pagnini2025,Pagnini2025b}. Black points show \odwarf{} members with membership probability $>70\%$. Blue points represent the Nephele stars assuming the same probability floor, $P_{\rm Nephele}>70\%$. The red dots are the Nephele stars in common with \odwarf{}.}
    \label{fig:nephele}
\end{figure}

In this context, GE and Sequoia+Thamnos are overdensities in the $E$–$L_Z$ space of such a massive dwarf galaxy, which can be disentangled using the data and tools we have today. While the more diffuse distribution spanning the entire $E$–$L_Z$ space is more difficult to obtain, since those stars are more well phased and chemically mixed with the MW field population. One successful attempt to recover this distribution was made by \citet{Pagnini2025,Pagnini2025b}. The so-called \textit{Nephele} debris were identified using a pipeline based on APOGEE and \textit{Gaia} information, which allowed the identification of debris everywhere in the Galaxy and in the $E$–$L_Z$ space. \autoref{fig:nephele} shows that, as expected from the above mentioned literature, the Nephele debris possesses all kinds of orbits (blue dots) populating the entire diamond diagram, including some stars in common (red dots) with our \odwarf{} sample (black dots).

{Nevertheless, even though simulations show that the debris populate the whole $E-L_Z$ space, it remains unclear why we only observe a bimodality in the action space. The AURIGA simulation employed in \citet{Skuladottir2025} shows that the perigalactic approximations (NSC motion) follow a clear path along the retrograde side of the $E-L_Z$ space, ending up in a $L_Z \sim 0.0$ orbit. Translating this into our scenario, the action and $E-L_Z$ space should present a more uniform distribution. Instead, we observe a change of orbital type, being more circular when the dwarf starts to fall into the MW, becoming more radial, and ending up on a circular orbit again. In this case, it is still unclear whether GE is part of the \odwarf{}. In the following analysis, therefore, we will also consider the scenario without GE.}

\subsection{Dwarf galaxy-like stellar population}\label{sec:dg_pop}

\begin{figure}[hbt!]
    \centering
    \includegraphics[width=0.99\linewidth]{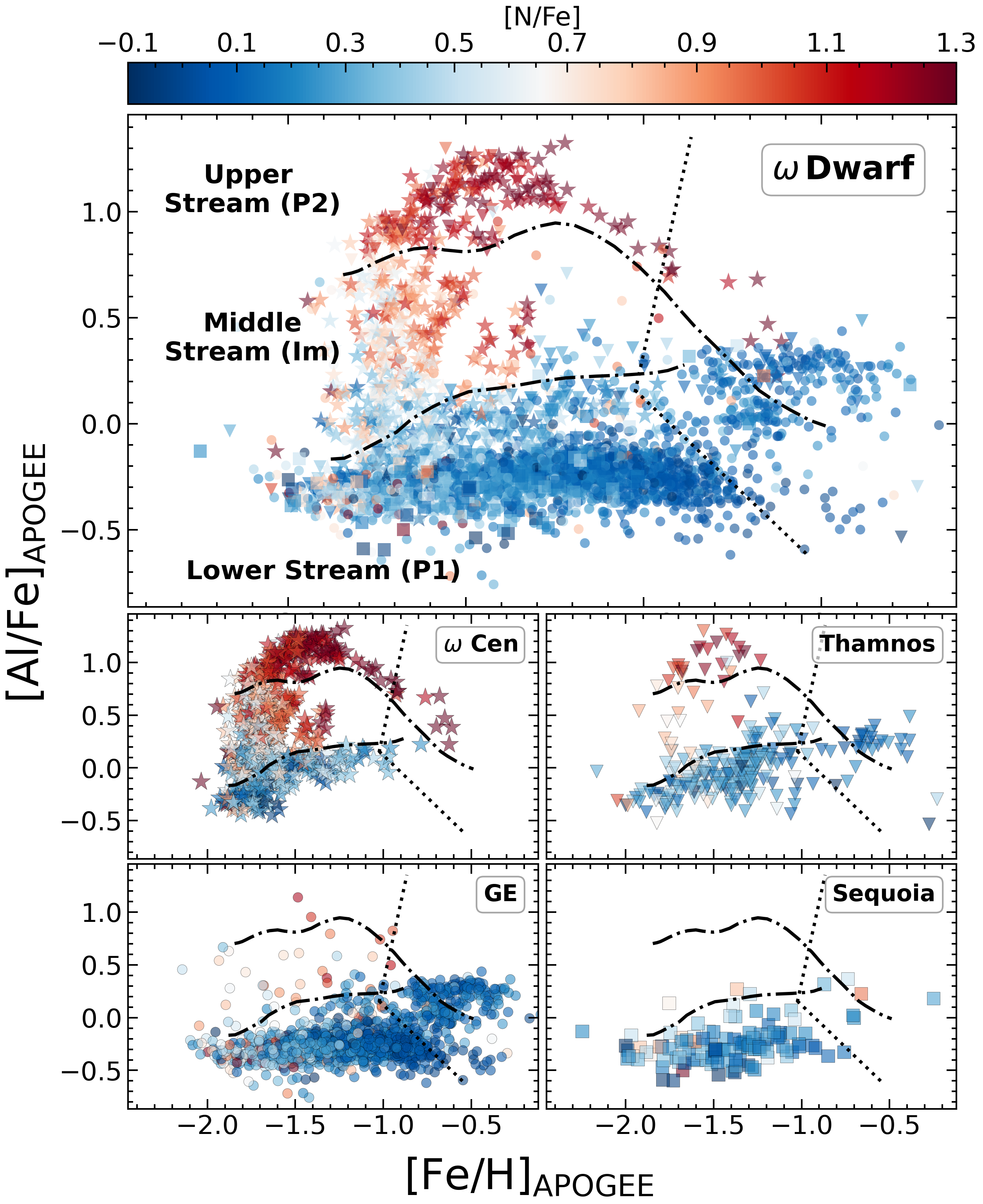}
    \caption{Stellar populations across \odwarf{} in the [Al/Fe]--[Fe/H] plane for APOGEE. The upper panel represents the entire sample of \odwarf{} members, whose symbols and colours follow the same definition as in \autoref{fig:target_selection}. Each \odwarf{} substructure is individually displayed in the bottom panels. The lines show the limits for the lower (P1) and upper (P2) streams as defined by \cite{Dondoglio2025}, while the dotted lines show the limits for the unknown population.}
    \label{fig:al_fe_apogee}
\end{figure}

During the merger, star formation can either be quenched or triggered \citep{Ernandes2024,Skuladottir2025}. If new star formation is initiated using up the remaining gas, the new stellar population will show some chemical peculiarities due to the enriched composition of the gas \citep{Matteucci2021}. In \autoref{fig:al_fe_apogee} we show the distribution of APOGEE [Al/Fe] as a function of [Fe/H] for \odwarf{} (upper panel) and its individual components (bottom panels), colour-coded by [N/Fe] (see \autoref{fig:al_fe_galah} for the case of GALAH). The evident correlation between [Al/Fe] and [N/Fe] \citep[also observed by ][]{Meszaros2021} leads to the same interpretation as in \cite{Dondoglio2025}. They interpreted this chemical plane as a chemical counterpart of the chromosome diagram \citep{Milone2017, Milone2022}, where the vertical axis reflects the difference in nitrogen abundance, while a difference in metallicity and helium mass fraction (Y) changes the horizontal axis \citep{Milone2018}.

\ocen{} presents the most complete set of populations in \odwarf{} as it is the surviving nucleus and continues evolving its stellar populations over time. The most comprehensive study aimed at understanding the formation history of \ocen{} has been conducted by the oMEGACat project\footnote{\url{https://omegacatalog.github.io/}}. The project consists of the most up-to-date catalogue of \ocen{} combining spectroscopic information \citep[][paper I]{Nitschai2023} for more than 300,000 stars using the Multi-Unit Spectroscopic Explorer (MUSE), and HST photometry and proper motions in 7 bands for $\sim1.4$ million stars \citep[][paper II]{Haeberle2024} for stars within the half-light radius ($\sim5'$). The high-precision astrometry of oMEGACat data allowed for the detection of fast-moving stars in the centre of \ocen{}, indicating the presence of an intermediate-mass black hole \citep{Haeberle2024Nature} and to obtain the most precise kinematic distance to \ocen{} of $5.494\pm0.061$ kpc \citep[][paper VI]{Haeberle2025}. The combined spectrophotometric analysis using the metallicity values from MUSE concluded that the inner region (within the half-light radius) of \ocen{} is completely spatially-mixed, showing no gradients \citep[][paper III]{Nitschai2024}. %The same work has also shown that the chromosome diagram of \ocen{} is a key tool for separating the different subpopulations based on chemical abundances. 
A further study based on the Na I D absorption lines in the MUSE spectra found no evidence of the presence of intracluster gas \citep[][paper VII]{Wang2025}.

The serial work presented in \citet[][paper IV]{Clontz2024}, \citet[][paper V]{Clontz2025}, \citep[][paper VIII]{Clontz2026}, and more recently \citet[][Paper IX]{Wang2026}, compile the most detailed analysis concerning \ocen{} stellar populations. Combining the ages from subgiant-branch stars \citep[SGBs][]{Clontz2024} with the identified 14 subpopulations in \ocen{} \citep[][]{Clontz2026}, the P2 populations were found to be at least 1 Gyr younger than the P1 populations and extremely He-enhanced \citep[][]{Clontz2025}. \cite{Wang2026} have shown additionally that P2 populations are systematically Na-richer than P1 and present the most extreme values of [Ba/Fe] ($\sim 1.0$). In parallel to oMEGACat studies, \cite{Mason2025} and \cite{Dondoglio2025} have also attempted to explain the complex chemical pattern of \ocen{} subpopulations. However, even though some of the proposed scenarios can explain certain chemical patterns, none fully accounts for the entire history. Nevertheless, the consensus is that the P2 population did form anyway after P1 \citep[see ][for a summary of each proposed scenario]{Wang2026}.

In the \odwarf{} scenario, if the P2 population, which means the Al-N-Ba-rich population, formed later, it might be confined to the inner region of the dwarf galaxy, since the outer parts are stripped first, before the formation of P2. \autoref{fig:al_fe_apogee} supports this hypothesis. While \ocen{} presents a large number of P2 stars, Sequoia, which we propose to be part of the outskirts, does not have P2 stars. Interestingly, Thamnos has a high fraction of P2 stars, indicating that it is more likely to have undergone a more recent stripping (relative to the infall time). Moreover, the Thamnos [Al/Fe]--[Fe/H] distribution resembles the \ocen{} one. GE shows a high density of low-Al, P1 stars, with a small fraction of Im stars.

\begin{figure*}[hbt!]
    \centering
    \includegraphics[width=0.99\linewidth]{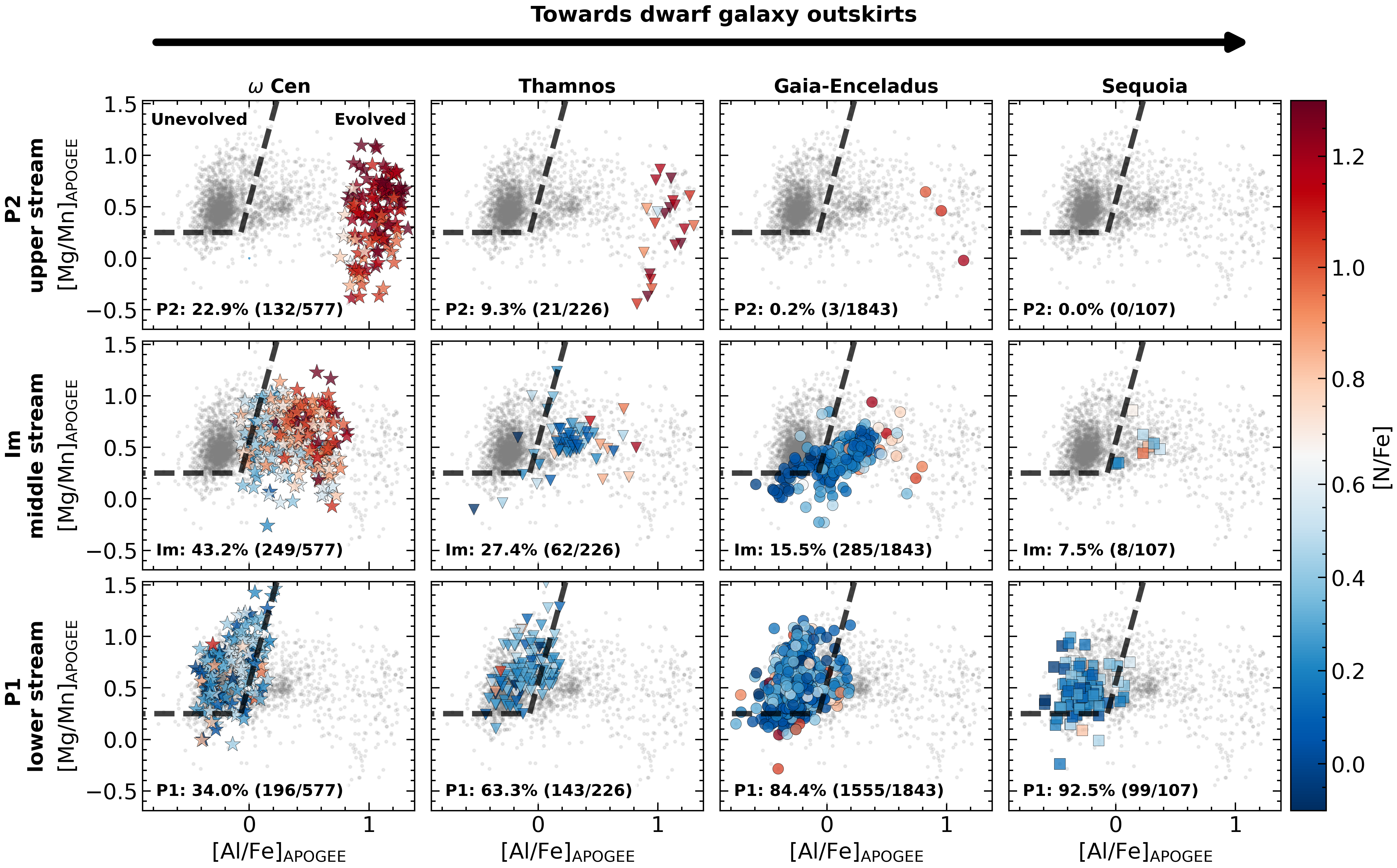}
    \caption{ Identification of \odwarf{} evolved and unevolved populations through the [Mg/Mn]--[Al/Fe] plane for APOGEE. The components of \odwarf{} are displayed in the columns: \ocen{} on the left, Thamnos in the middle left, GE in the middle right, and Sequoia on the right. The top row represents the P2 population / upper stream, the middle stream in the central row, and the P1 population / lower stream in the bottom row. The color code indicates [N/Fe] values. The fraction (in percentage and absolute number) of stars in each population is indicated in the bottom left in each panel. Only stars with measured [Al/Fe] and [Mg/Mn] abundances were considered.}
    \label{fig:mgmn_apogee}
\end{figure*}

The same pattern is observed with more detail in \autoref{fig:mgmn_apogee} for APOGEE data (see \autoref{fig:mgmn_galah} for the GALAH case). The [Mg/Mn]--[Al/Fe] plane is an efficient diagnostic for distinguishing the evolved population from the unevolved one \citep{Hawkins2015, Das2020, Lane2023} because the ratios represent different nucleosynthesis mechanisms. Mg and Al are predominantly produced in core-collapse SNe (CCSN), and Mn and Fe have significant contributions from SNIa \citep{Kobayashi2020}. The relative timescales and yields of these enrichment channels depend on star-formation efficiency and the initial mass function (IMF), leading to distinct sequences that can differentiate stellar populations \citep{Vasini2024}. The unevolved population region was defined empirically as \citep{Hawkins2015}:
\begin{align}
\mathrm{[Mg/Mn]} > 0.25\quad {\rm and} \quad \mathrm{[Mg/Mn]} > 5 \times \mathrm{[Al/Fe]} + 0.5. \label{eq:chem_region}
\end{align}

In \autoref{fig:mgmn_apogee}, the columns represent each \odwarf{} component individually, with the separation into the P1, Im, and P2 populations shown in the bottom, middle, and top rows, respectively. The P2 population (upper panels) disappears towards the outskirts, while the P1 fraction increases. Moreover, the P1 population clearly falls into the unevolved region, confirming its primordial origin. Here, it is interesting to note that the Thamnos stars again follow the same pattern as the \ocen{} ones. The fact that P2 stars are not populating the outskirts may suggest that P2 stars formed from a source of self-enrichment, such as asymptotic giant branch stars (AGBs) feedback, as suggested by \cite{Dondoglio2025}. In the following analysis, we focus on the P1 population, since the P2 stars are more than 1 Gyr younger and likely formed during or after the merger between \odwarf{} and the MW. The origin of the P2 population will be further investigated in another work (Clontz et al. in preparation).

\subsection{Dwarf galaxy-like enrichment}

\cite{Leaman2012} analysed spectroscopic data from numerous Local Group dwarf galaxies and star clusters, demonstrating that these objects occupy distinct sequences in the global metallicity intrinsic variance versus mean global metallicity diagram ($\sigma(Z)^2$ versus $\bar{Z}$). This separation arises fundamentally from their different star formation timescales and enrichment processes. Dwarf galaxies, with their extended star formation histories, undergo multiple, spatially inhomogeneous enrichment events from SNe, leading to a large intrinsic spread in metallicity around a given mean. This self-enrichment signature defines a characteristic sequence for dwarf galaxies in the diagram.

To verify if a sample of stellar debris is consistent with a dwarf galaxy origin, one would then position it in the $\sigma(Z)^2$ versus $\bar{Z}$ diagram. If the debris originates from a disrupted dwarf galaxy, its locus should fall on or near the well-defined sequence occupied by known dwarf galaxies. In contrast, a sample from a star cluster would exhibit a much smaller metallicity spread for its mean metallicity, placing it on a separate sequence offset from the dwarf galaxy trend. We plotted the different substructures we analysed on the $\sigma(Z)^2$ versus $\bar{Z}$ diagram, deriving the intrinsic variance and mean metallicity using the same approach as in \cite{Leaman2012}. 

\begin{figure}[hbt!]
    \centering
    \includegraphics[width=0.99\linewidth]{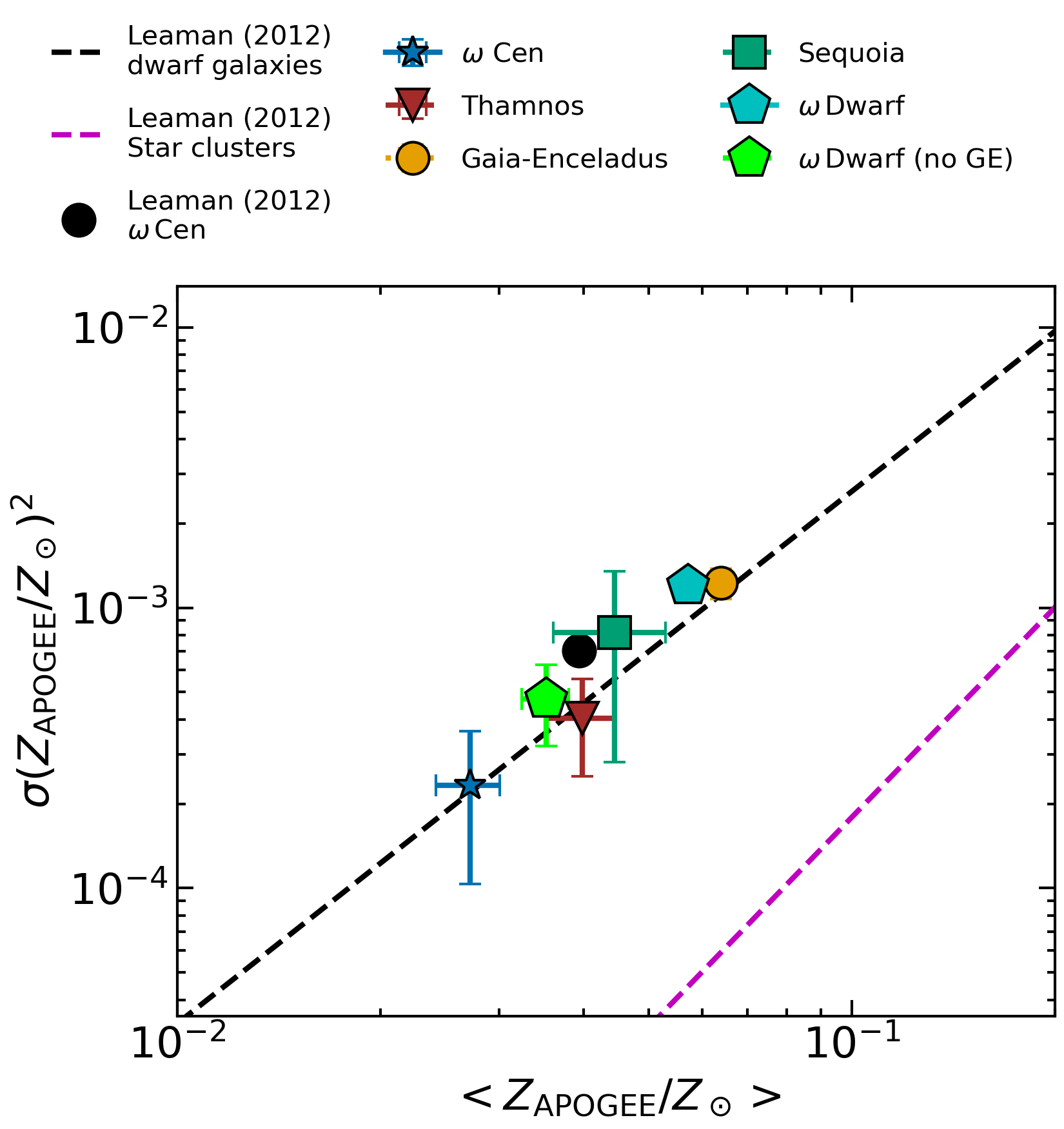}
    \caption{Intrinsic APOGEE metallicity variance vs. mean metallicity for \odwarf{} P1 population (lower stream). The symbols represent the substructures individually, and the black dot shows the position of \ocen{} derived by Leaman (2012) assuming all populations. The black-dashed line shows the linear regression for dwarf galaxies, while the magenta one is for star clusters. The cyan and lime pentagons represent the mean loci of \odwarf{} considering or not GE stars, respectively. This shows that, considering both the whole \odwarf{} or its components individually, the scenario is consistent with a dwarf galaxy chemical enrichment. The error bars represent $3\times$ the standard deviation.}
    \label{fig:sigmaZ-Z}
\end{figure}

In \autoref{fig:sigmaZ-Z} we display the linear relation for dwarf galaxies found by \cite{Leaman2012} as well as their position for \ocen{}. We used a jackknife method to estimate errors. However, as large as the sample is, the method returns smaller errors; we show $3\sigma$ error bars in \autoref{fig:sigmaZ-Z} to make them visible. %This is why Sequoia shows a larger error bar than the others, whereas for GE, the error bar is not visible. 
Since we consider only the lower stream (P1) population, we find a different position for \ocen{} compared to their determination. The dwarf galaxy nature of each substructure is evidenced by the match with the dashed-black line, while the possible connection among them is supported by their linear correlation. The sequence in metallicity is also compatible with the scenario proposed in \autoref{fig:scenario}, showing the metallicity increasing toward the center of the dwarf galaxy. However, GE, which is proposed to be the intermediate region, is more metal-rich than Sequoia, which, in \autoref{fig:scenario}, would form the outer regions of the dwarf galaxy. This will be discussed in \autoref{sec:discussion}.

\subsection{Dwarf galaxy-like Metallicity Gradient} \label{sec:energy_proxy}

\cite{Carrillo2026} showed that orbital energy can retain information about the pre-merger metallicity gradient of a dwarf galaxy. They found that, although the radial metallicity gradient differs considerably between pre- and post-merger ($\sim 93\%$), the metallicity pre/post-merger gradient based on orbital energy differs in a wider range ($17-70\%$). Motivated by this result, we assume that the present-day orbital energy of an accreted star encodes, in a statistical sense, its original location within the disrupted progenitor. We therefore use present-day orbital energy to define a normalised \textit{pre-merger radius} of \odwarf{}, with the highest energy (less bound) corresponding to 1, and lower energy (more bound) corresponding to 0. This definition enables us to examine radial gradients within the original dwarf galaxy and to test the consistency of our proposed scenario.

\begin{figure*}
    \centering
    \includegraphics[width=0.99\linewidth]{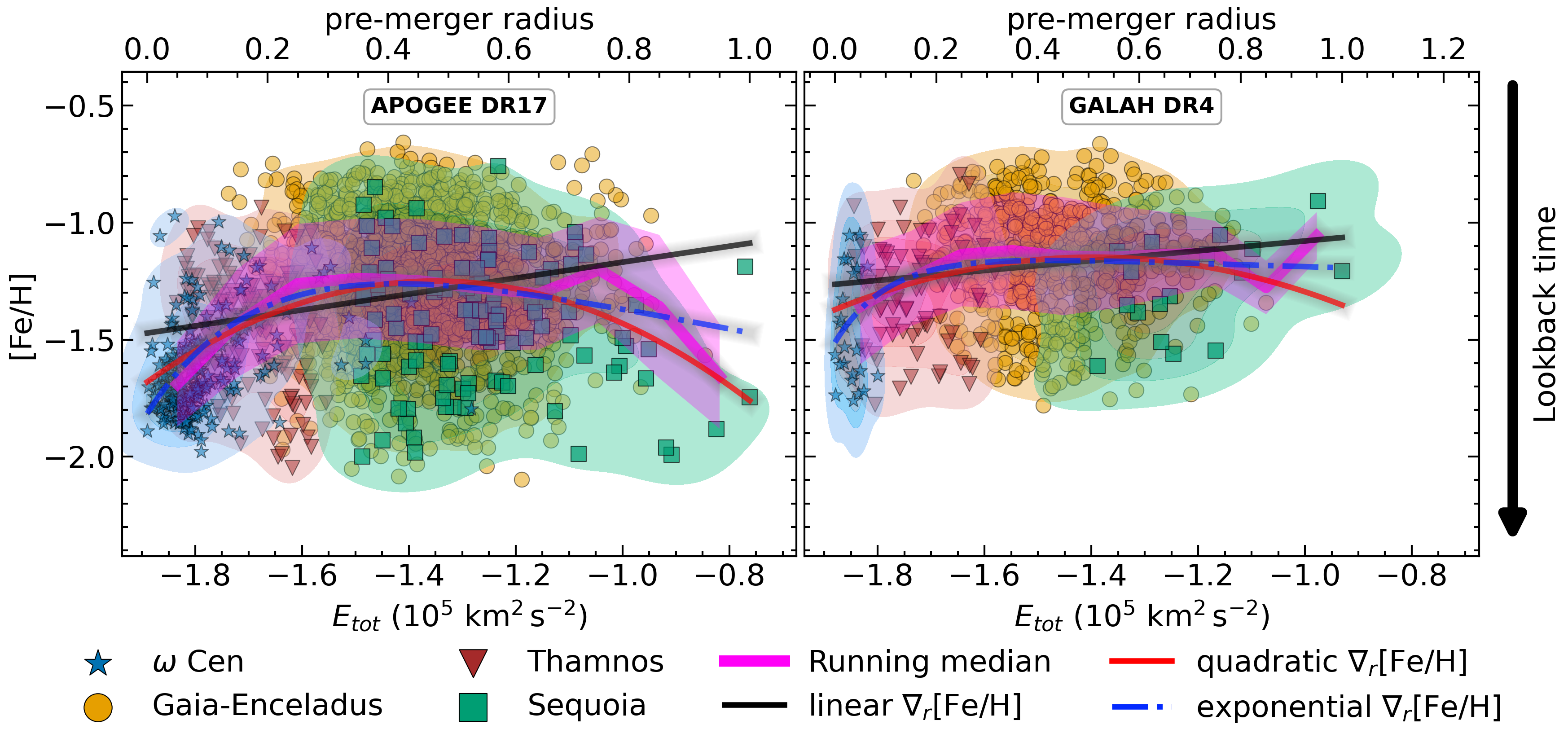}
    \caption{Metallicity as a function of total orbital energy, $E_{\rm tot}$, used as a proxy for pre-merger radius, for stars in APOGEE~DR17 (left) and GALAH~DR4 (right). The symbols and colours are the same as in \autoref{fig:target_selection}%: \ocen{} as blue stars, Thamnos as red triangles, GE as yellow dots, and Sequoia as green squares
    . The solid and dashed curves show different functional forms fitted to the combined samples: a linear relation (black), a quadratic model (red), and an exponential model (blue dashed). The magenta solid line is the running median curve. Even though the metallicity scatter is large along the pre-merger radius, the median running median curve evidences an inverted U-shaped gradient.  }
    \label{fig:met_grad}
\end{figure*}

The metallicity gradient in \autoref{fig:met_grad} shows an inverted U-shaped structure with energy, suggesting lower metallicity stars in both the centre of \odwarf{} and in its outskirts. To better display this feature, we defined a running median curve constructed by binning the energy axis (pre-merger radius) and calculating the median for each bin. The number of bins follows the definition \texttt{max($11; 2\times N^{1/3}$)}, where $N$ is the sample size. We show that a non-linear, exponential (blue dashed line) or quadratic (red solid line) relation is more likely to follow the running median curve (magenta solid line).

There are several ways that this U-shaped metallicity gradient could be understood in the context of typical scenarios of galaxy formation. Similar gradients are typical for nucleated dwarf galaxies  \citep[$\lesssim 10^9$ M$\odot$;][]{Fahrion2021, Fahrion2022}. One example, FCC119 in \cite{Fahrion2021}, provides an instructive comparison: its NSC mass, $\log(M_{\rm NSC}/M_\odot)=6.81$, is comparable to that of \ocen; its host-galaxy mass, $\log(M_\star/M_\odot)=9.0$, lies within the range typically inferred for a GE-mass progenitor; and its stellar population is broadly consistent with that expected for the \odwarf{} at the time of merger. A straightforward explanation for FCC119 is that the NSC could have formed via the merger of inspiraling metal-poor GCs \citep{Tremaine1975, Kacharov2018, Neumayer2020}. 

The \odwarf{} metallicity gradient (\autoref{fig:met_grad}), from a midpoint to the outskirts, displays a negative gradient more commonly associated with an outside-in formation \citep[e.g.][]{Leaman2013, Kacharov2017}, also a hallmark of strong outflows in low-mass systems, which effectively suppress metal enrichment by ejecting processed material into the circumgalactic medium \citep[e.g.][]{Larson1974, Kirby2013}. In contrast, the positive gradient observed in \odwarf{} from the centre to a midpoint points to a formation via merger of inspiral GCs, as for FCC119. Nevertheless, in situ parallel formation from concentrated material is not discarded \citep{Guillard2016}. 

In a more realistic treatment of the pre-merger radius, the inner positive gradient may originate from a small range of radii in the original dwarf. This material could have been harder to disrupt and thus spread out more in energy since it was sinking in the MW potential over a longer period of time, whereas the outer regions wee stripped earlier. This would make the \odwarf{} chemical enrichment clearer. Nevertheless, in the following section, we discuss the origin of the metallicity gradient inferred for the \odwarf{}, supported by additional chemical abundance diagnostics.

\section{Adding details: \\Abundance Variations within \odwarf{}}\label{sec:discussion}

In the following subsections, we discuss different enrichment mechanisms based on this gas inflow/outflow formation scenario.

\subsection{The $\alpha$ elements across \odwarf{}}

One key point of the metallicity gradient in \autoref{fig:met_grad} is that the inner part (\ocen{}+Thamnos) corresponds to the formation of the NSC via the inspiral of GCs. Another way to compare the evolution of the inner (\ocen{}+Thamnos) and outer (GE+Sequoia) regions is by using the [$\alpha$/Fe] ratio. We assume [$\alpha$/Fe] as the mean value of O, Mg, Si, and Ca that are predominantly synthesised and released on short timescales ($\lesssim 10\, Myr$) by CCSN. In contrast, the production of iron mainly from SNIa is significantly delayed \citep[$\gtrsim 10^8$–$10^9$ yr,][]{Tinsley1979,Matteucci1990,McWilliam1997,Sneden2008}.

\begin{figure}[hb!]
    \centering
    \includegraphics[width=0.99\linewidth]{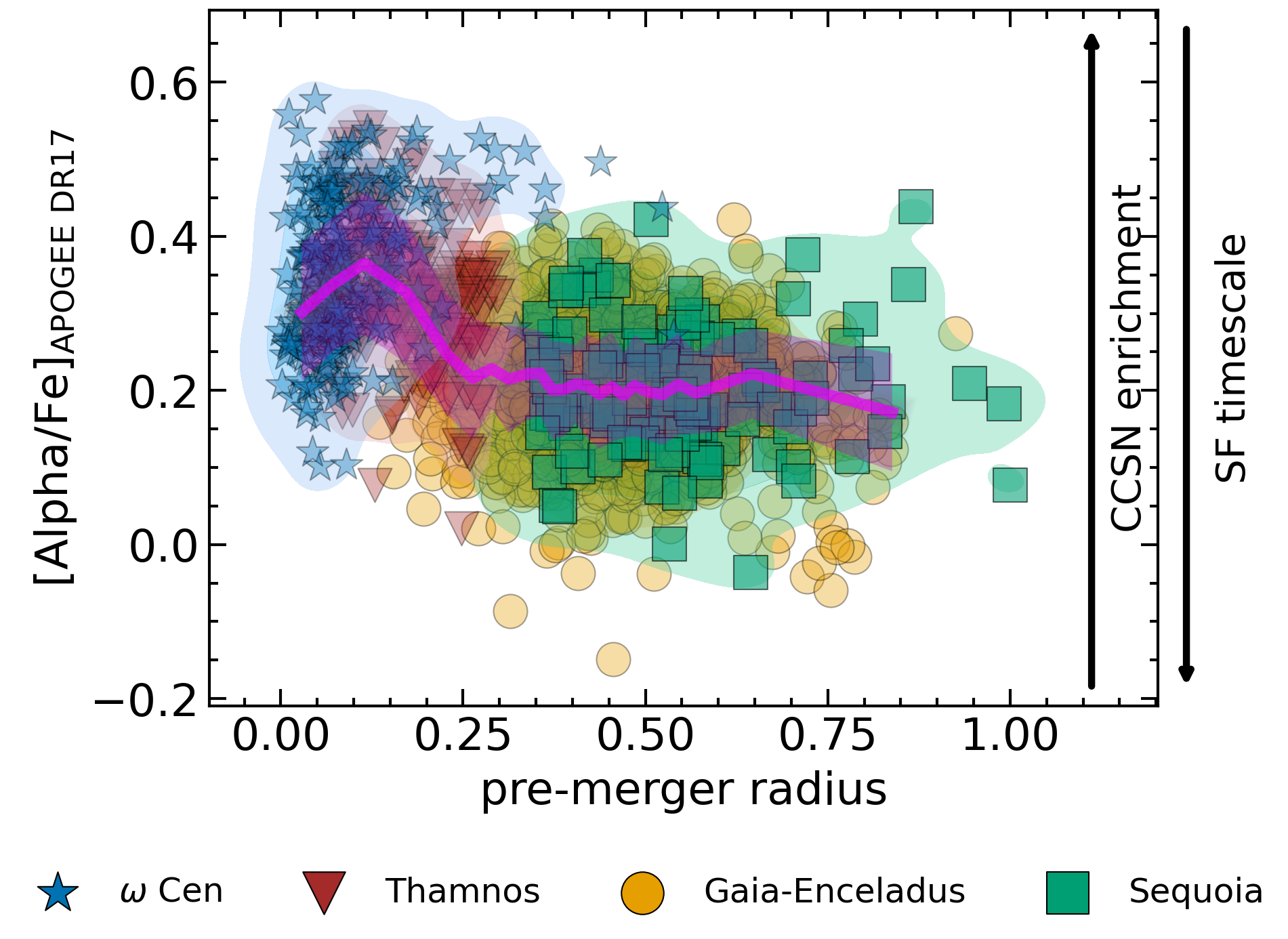}
    \caption{ Abundance of $\alpha$ elements across \odwarf{}. The [$\alpha$/Fe] ratio is the mean of O, Mg, Si, and Ca abundances. The symbols and colours follow the same code as in \autoref{fig:target_selection}%: \ocen{} as blue stars, Thamnos as red triangles, GE as yellow dots, and Sequoia as green squares
    . The magenta line is the running median of the distribution. The y-axis can be used as a proxy for both CCSN enrichment and the SF timescale. Then, the inner part (\ocen{}+Thamnos) seems to have experienced a faster enrichment by CCSN than the outskirts (GE+Sequoia).
    }
    \label{fig:alpha_grad}
\end{figure}

The gradient of [$\alpha$/Fe] in \autoref{fig:alpha_grad} shows that the inner region (\ocen{}+Thamnos) is more enhanced than the outskirts (GE+Sequoia). Since the ratio [$\alpha$/Fe] measures the relative contribution of CCSN versus SNIa, higher values indicate a greater contribution from CCSN. This means that the formation of the inner stellar population occurred before the SNIa contribution became more relevant, indicating a high star formation efficiency (SFE) and, consequently, a short star formation timescale ($\tau_{\rm SF}$) \citep{Matteucci2021}. The $\Delta$[$\alpha$/Fe] of $\sim 0.15$ dex between the inner region and the outskirts indicates that the formation of \ocen{}+Thamnos happened in a more efficient and fast star formation burst, typical of GC formation. GCs usually form in a highly efficient (SFE $\sim 5-20$ Gyr$^{-1}$) and short ($\tau_{\rm SF} \sim 10-100$ Myr) star formation episode \citep[e.g.][]{Romano2007,Romano2023, Matteucci2021}. In contrast, dwarf galaxies form stars less efficiently (SFE $\sim 0.01-1$ Gyr$^{-1}$) and take longer time to quench ($\tau_{\rm SF} \sim 2-8$ Gyr) \citep[e.g.][]{Tolstoy2009, Romano2005, Romano2015}. Therefore, the gradient in  [$\alpha$/Fe] also supports the scenario where the inner region (\ocen{}+Thamnos) forms via inspiral of GCs while the outskirts evolves in a more dwarf galaxy-like scenario.

\begin{figure}[ht!]
    \centering
    \includegraphics[width=0.99\linewidth]{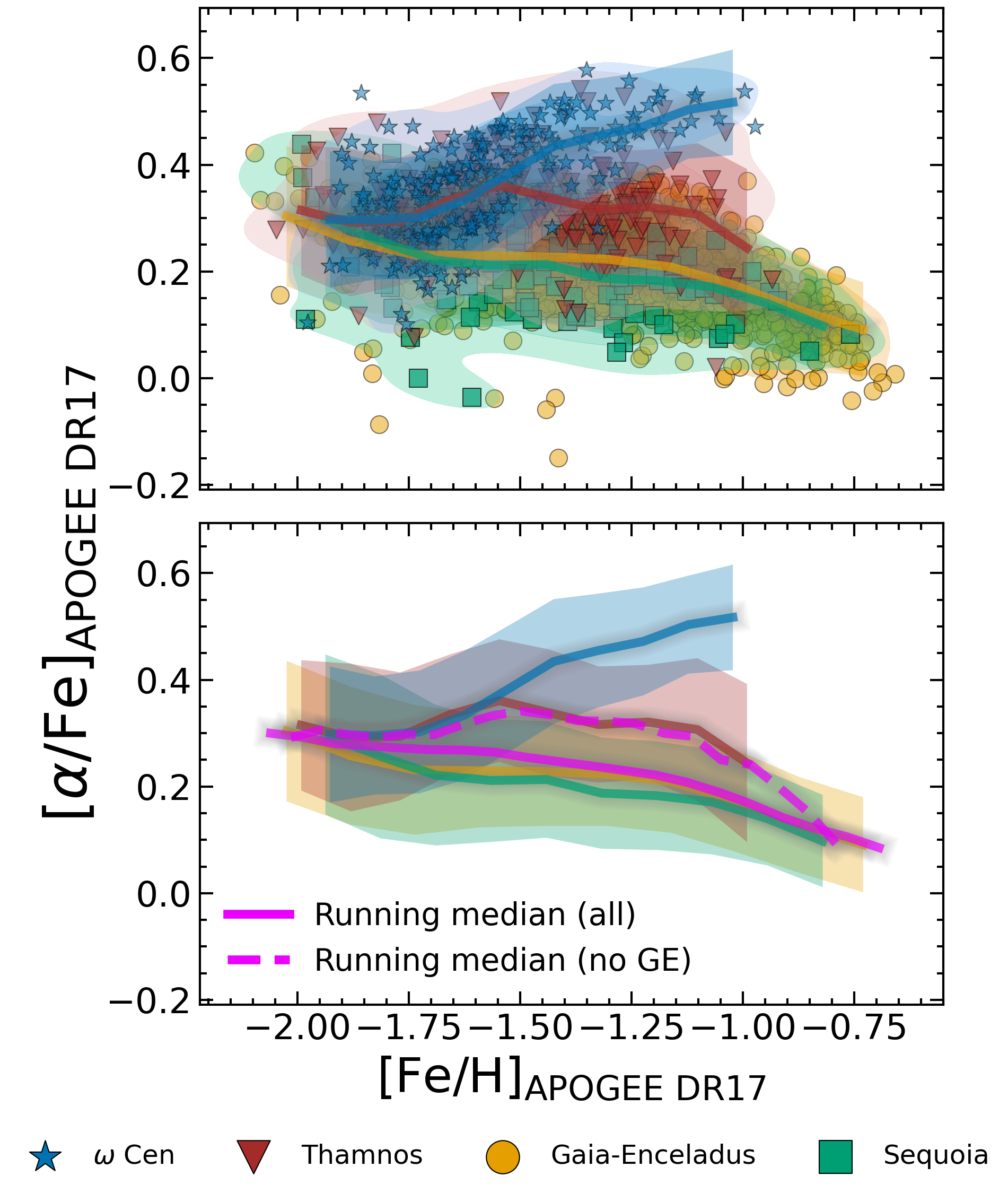}
    \caption{ The $\mathrm{[\alpha/Fe]}$ ratio as a function of metallicity of \odwarf{} using APOGEE. The dotted line and the shaded region indicate the position of the metallicity knee and the confidence interval, respectively. The symbols and colours are the same as in \autoref{fig:target_selection}. The difference in the magenta lines in the bottom panel points to a difference in the total mass of \odwarf{} considering or not GE (solid and dashed lines respectively).
    }
    \label{fig:alpha_fe}
\end{figure}

The [$\alpha$/Fe] as a function of metallicity in \autoref{fig:alpha_fe} shows that each \odwarf{} component has a different slope of the distribution. The upper panel of the figure shows the stellar distribution with running median lines overplotted following the same colour code as for the symbols: \ocen{} (blue), Thamnos (red), GE (yellow), and Sequoia (green). While GE and Sequoia have almost equal running medians, Thamnos shows a more constant distribution of [$\alpha$/Fe] around $0.3$, and for \ocen{} [$\alpha$/Fe] increases with metallicity. Essentially, the slope becomes positive towards the inner regions. The GC inspiral scenario for the formation of the inner region emerges again via the metal-rich, $\alpha$-enhanced tail for \ocen{}, which can be better explained by assuming different-metallicity GCs \citep[e.g.][]{Guillard2016,Clontz2026,Wang2026}. 

The bottom panel of \autoref{fig:alpha_fe} shows the running median lines, along with those for the entire sample (magenta-solid) and the case without considering GE (magenta-dashed). Avoiding GE for the \odwarf{} running median makes the curve more $\alpha$-enhanced and slightly offset towards metal-poor, since the bulk of metal-rich stars are in the GE distribution. This diagonal offset implies some differences in the global evolution of \odwarf{}. The higher plateau in [$\alpha$/Fe] can be understood as an increase in the global SFE and fewer SNIa events per mass of the dwarf galaxy. At the same time, a stronger outflow shifts the upper metallicity limit to lower values \citep[see][for more details on chemical evolution model parameters]{Matteucci2021}. 

Another consequence of avoiding GE for the \ocen{} scenario is the calculation of stellar mass. The knee in the [$\alpha$/Fe]--[Fe/H] plane marks the metallicity at which the enrichment from delayed SNIa begins to dominate over that of CCSN, causing the [$\alpha$/Fe] ratio to drop. Because the timing of this transition depends on the galaxy’s star‑formation efficiency and gas‑retention ability -- both correlated with its total mass -- the metallicity knee, [Fe/H]$_{\rm knee}$, is empirically linked to the stellar mass of the galaxy \citep[e.g.][]{Matteucci2014,Matteucci2021}. A more metal-poor knee represents a less massive dwarf galaxy \citep[e.g.][]{Hansen2018, Mason2024}. In that sense, in a qualitative analysis, avoiding GE, the mass of \odwarf{} tends to be lower than that estimated for a GE-like dwarf \citep[the latest $4.49^{+5.69}_{-2.36}\times10^9\, M_\odot$,][see also \citet{Helmi2018}]{Massari2026}. To better constrain the \odwarf{} mass, a chemical evolution modelling-based analysis is needed (Souza et al., in preparation).

% -----------------------------------------------------------------------------
\subsection{r-process elements across \odwarf{}}

The rapid neutron-capture (r-process) is responsible for creating approximately half of the elements heavier than iron, including Eu, which is an almost pure r-process tracer, and a small fraction of Ba, which is predominantly produced in AGB stars via the slow neutron capture (s-process) mechanism \citep{Kobayashi2020,Prantzos2020}. Identifying the astrophysical sites of the r-process -- such as CCSN, neutron star mergers \citep[NSMs; ][]{Watson2019} or Magneto-rotational supernovae \citep[MRSN][]{Nishimura2019} -- remains a central challenge \citep{Cote2019,Molero2023, Cavallo2023}. In low-mass dwarf galaxies, the stochastic nature of star formation and enrichment means these rare events can dominate the chemical signature of subsequent stellar generations \citep{Ji2016, Tsujimoto2017}.

To investigate the extent of r-process enrichment relative to instantaneous chemical enrichment by massive stars, we examine the abundance ratio of r-process to alpha elements. While Mg is widely used, we follow \citet{Monty2024} and use Si as an alpha element to avoid the potential Mg depletion in globular clusters. If Eu is also produced by an instantaneous enrichment, similarly to alpha-elements, the ratio [Eu/Si] is expected to stay flat with time. On the contrary, it would increase with time if it has a delayed source, such as NSM. In the latter case, [Eu/Si] of different populations would also be different, reflecting their star formation timescales. 

\begin{figure}[hbt!]
    \centering
    \includegraphics[width=\linewidth]{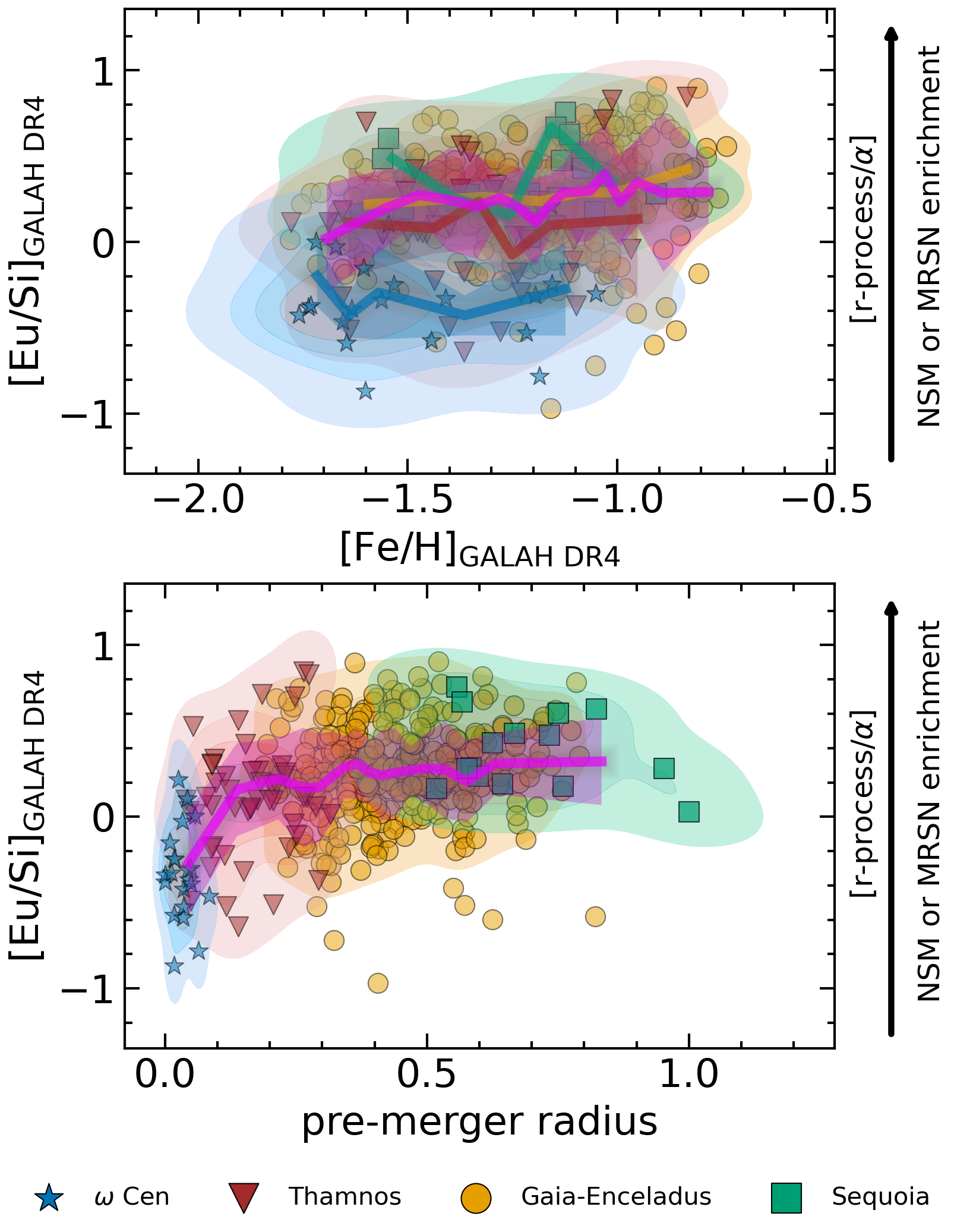}
    \caption{ NSM enrichment in \odwarf{} using GALAH. The top panel shows the abundance ratio [Eu/Si] as a function of metallicity, while the bottom panel shows it as a function of the pre-merger radius. The symbols and colours are the same as in \autoref{fig:target_selection}. The magenta line, representing the running median, shows that the outskirts of \odwarf{} is more r-enriched than the central part. }
    \label{fig:eu_si}
\end{figure}

In the top panel of \autoref{fig:eu_si} we show the [Eu/Si] as a function of [Fe/H] for \odwarf. In the central region, \ocen{} and Thamnos show no correlation between [Eu/Si] and metallicity (blue and red solid lines), indicating a enrichment process not from rare events (e.g. NSM or MRSN). In the outskirts, GE (yellow solid lines) shows a weak positive correlation with metallicity. For Sequoia (green solid line), it is not possible to infer a correlation due to low sample sizes. When observing the [Eu/Si] gradient (bottom panel of \autoref{fig:eu_si}), the enrichment becomes clearer. The outskirts present a plateau around [Eu/Si]$\sim0.25$ with a high dispersion.

\begin{figure}[hbt!]
    \centering
    \includegraphics[width=\linewidth]{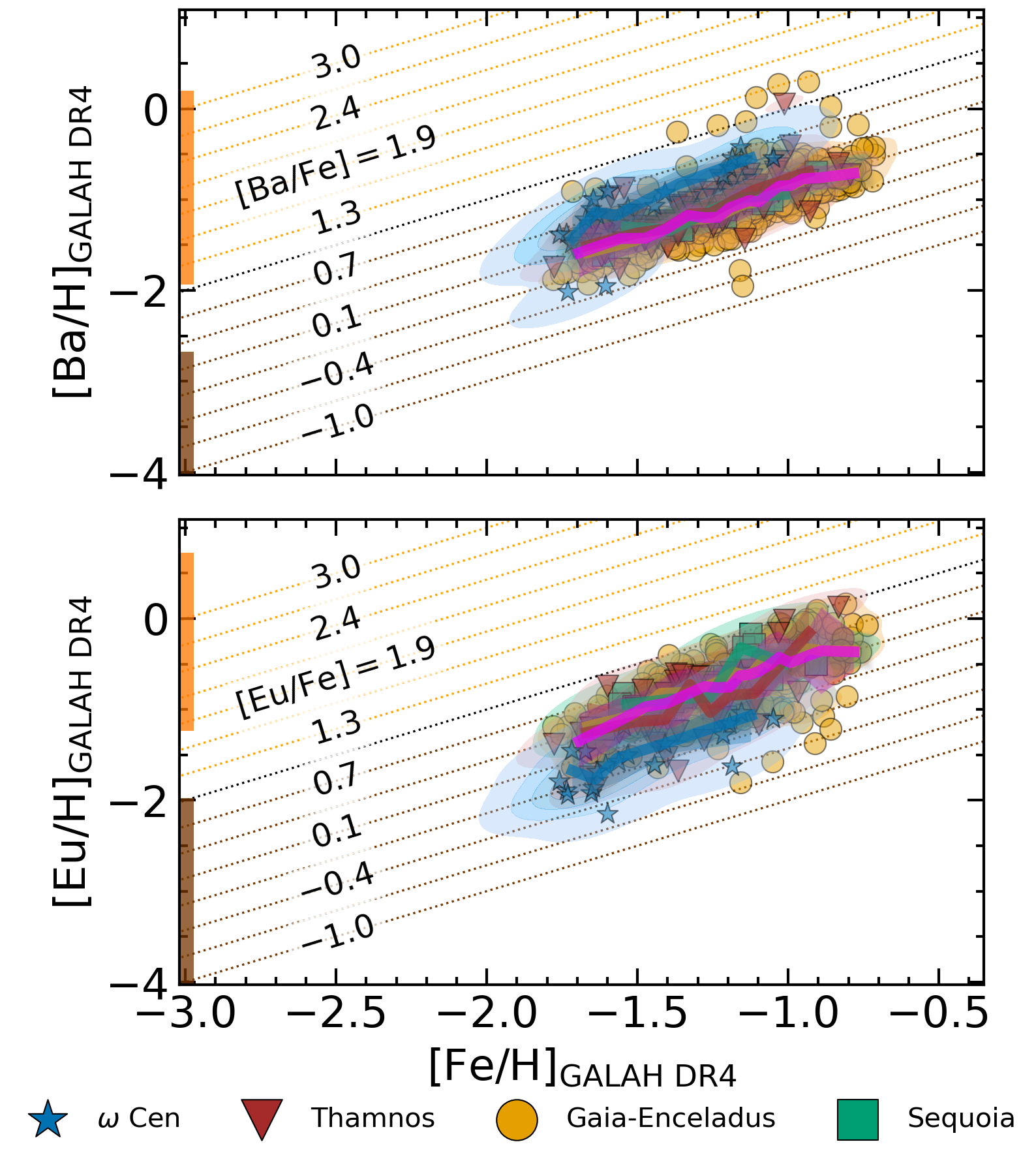}
    \caption{ The s- and r-process elements in \odwarf{} for GALAH. The [Ba/H] and [Eu/H] abundances as a function of metallicity follow the same colour and symbol color code as in \autoref{fig:target_selection}%: \ocen{} as blue stars, Thamnos as red triangles, GE as yellow dots, and Sequoia as green squares
    . The dotted diagonal lines indicate constant [Ba/Fe] (top) and [Eu/Fe] (bottom) values ranging from -0.8 to 1.3. The running median lines are colored according to their population, while the magenta curve shows the overall median trend of the sample. The orange and brown vertical bars represent the Ba and Eu ranges from which enrichment originated from NSM and CCSN, respectively. Following the constant [Ba/Fe] and [Eu/Fe] lines, no NSM enrichment is observed in \odwarf{}.}
    \label{fig:nsm_ccsn}
\end{figure}

In \autoref{fig:nsm_ccsn} we show the [Ba/H] (top panel) and [Eu/H] (bottom panel) as a function of metallicity. Following \cite{Ji2016}, we overplotted the lines of constant [Ba/Fe] and [Eu/Fe] (dotted lines), and the range of [(Ba or Eu)/H] abundances where it is expected for NSM (orange vertical bar) and CCSN (brown vertical bar). Seven stars in GE and one in Thamnos possess Ba abundances which could be interpreted as NSM enrichment. However, their Eu abundances, as well as those of the entire sample, are more consistent with an enrichment from CCSN. This result, therefore, indicates that the [Eu/Si] gradient of \autoref{fig:eu_si} (bottom panel) is likely to represent a higher r-process enrichment in the outskirts instead of a rare nucleosythesis channel (such as NSM or MRSN).

\subsection{s-process elements across \odwarf{}}

The heavy-element ratios across \odwarf{} indicate that the neutron-capture enrichment history is not spatially uniform, and that the inner and outer regions likely experienced different balances between delayed s-process production and prompt r-process events. In this context, as Ba and La (second s-process peak elements) mainly trace the heavy-s component, the ratios [Ba/Eu] and especially [La/Eu] are robust diagnostics of the relative s-/r-process contribution \citep[e.g.][]{Travaglio1999,Lanfranchi2005,Lanfranchi2008,Cescutti2006,Kobayashi2020}. 

\begin{figure*}[hbt!]
    \centering
    \includegraphics[width=0.99\linewidth]{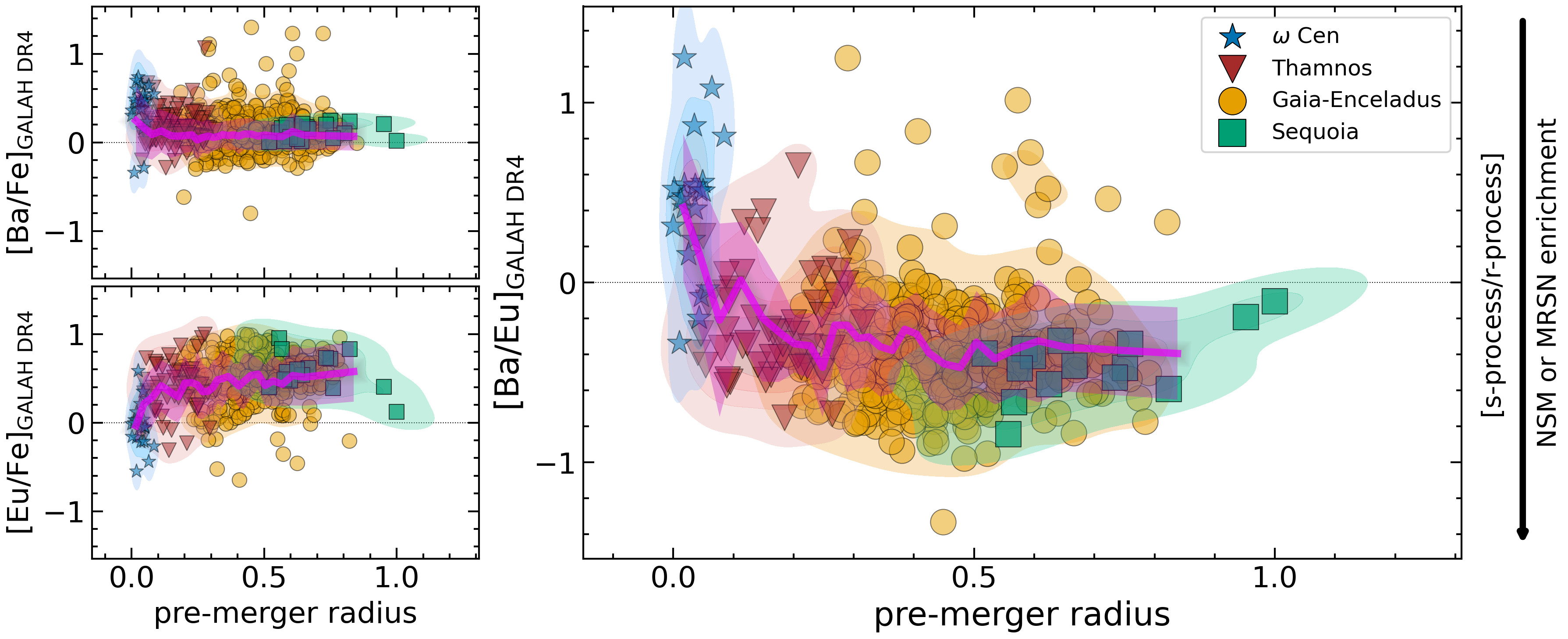}
    \caption{Heavy-element abundance trends as a function of pre-merger radius. Left panels: [Ba/Fe] (top left) and [Eu/Fe] (bottom left) versus pre-merger radius for \odwarf. The symbols and colours are the same as in \autoref{fig:target_selection}. Right panel: [Ba/Eu] versus pre-merger radius for the same samples. The magenta curve traces the median trend. Horizontal dotted lines mark the solar ratios. The vertical annotations indicate the dominant nucleosynthetic regimes, from s-process--dominated enrichment (AGB) at higher [Ba/Eu] to r-process--dominated enrichment (NSM or MRSN) at lower [Ba/Eu]. }
    \label{fig:eu_grad}
\end{figure*}

Figure~\ref{fig:eu_grad} shows that [Ba/Fe] remains nearly constant with pre-merger radius, with at most a weak decline toward the innermost regions. By contrast, [Eu/Fe] increases outward, such that the outer \odwarf{} populations associated with GE and Sequoia are more Eu-rich than the inner populations associated with \ocen{} and Thamnos. This combination naturally produces the radial decline in [Ba/Eu], with the outer regions characterized by [Ba/Eu]$\sim -0.5$, reflecting a strong r-process contribution. In contrast, the inner regions shift toward progressively larger [Ba/Eu], culminating in the extreme \ocen{} population with [Ba/Eu]$>0.8$. Such high [Ba/Eu] values strongly suggest a substantial enhancement of the heavy-s component relative to Eu, as expected if the central regions retained AGB ejecta more efficiently and experienced a more prolonged enrichment history than the outskirts \citep{Lanfranchi2005,Lanfranchi2008,Smith2000,Johnson2010}. 

\begin{figure}[ht!]
    \centering
    \includegraphics[width=0.99\linewidth]{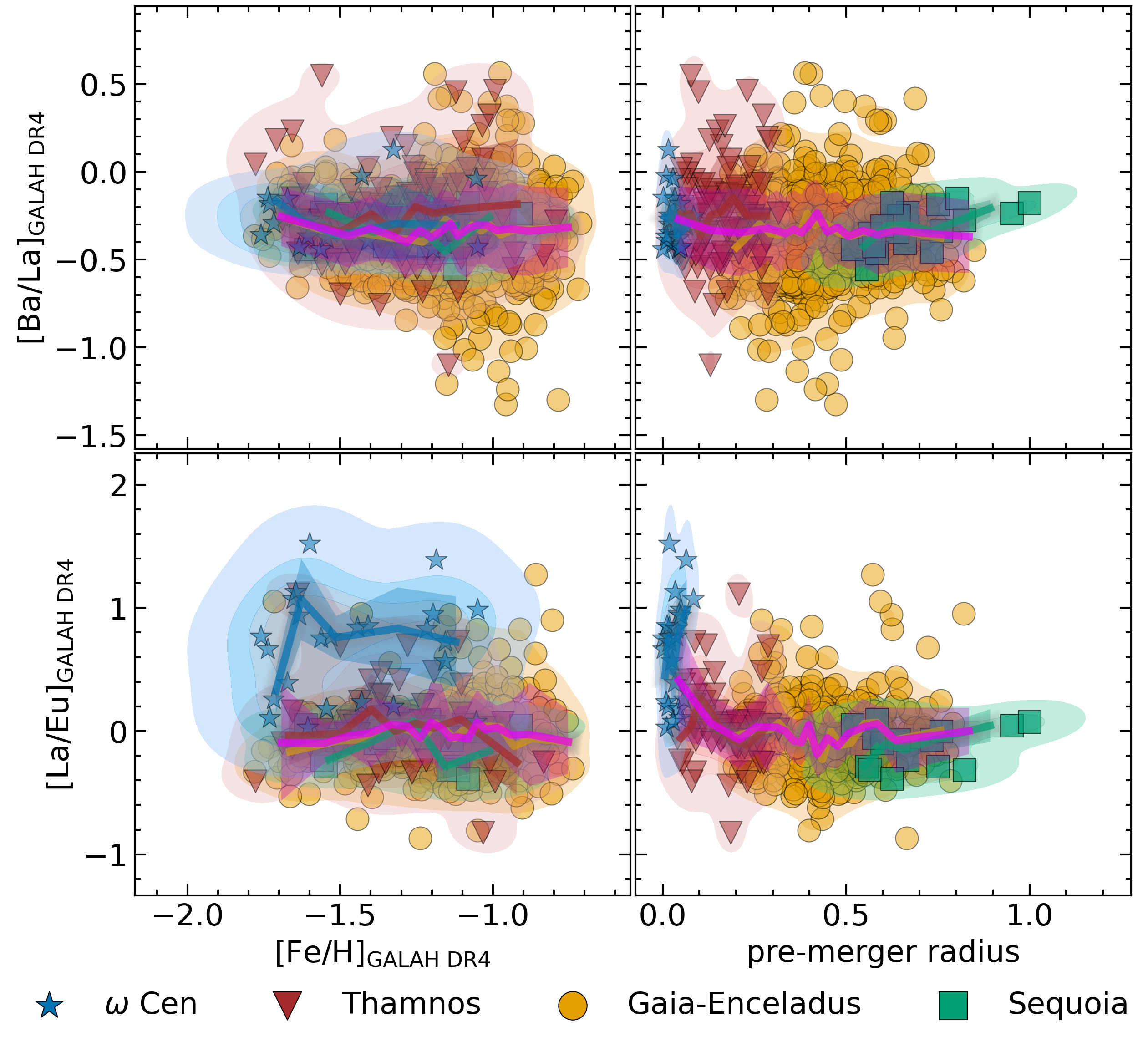}
    \caption{Neutron-capture abundance ratios across \odwarf{} using GALAH DR4. Top panels: [Ba/La] as a function of metallicity (left) and pre-merger radius (right). Bottom panels: [La/Eu] as a function of metallicity (left) and pre-merger radius (right). Symbols, lines, and colours identify the different \odwarf{} components as in \autoref{fig:target_selection}, and the magenta line shows the median trend. While [La/Eu] is highest in the inner regions, [Ba/La] is nearly constant, indicating that the centre of \odwarf{} is more strongly s-process enriched than the outskirts.}
    \label{fig:la_grad}
\end{figure}

The same behaviour is seen in Figure~\ref{fig:la_grad}. The ratio [Ba/La] is approximately constant at [Ba/La]$\sim -0.25$ over the full pre-merger radius range. This indicates that Ba and La largely track one another, and that the dominant variation is not internal to the heavy-s group itself. In contrast, [La/Eu] shows the same qualitative radial behaviour as [Ba/Eu], with the highest values concentrated toward the inner \odwarf{} regions. The agreement between the [Ba/Eu] and [La/Eu] gradients therefore strengthens the conclusion that the centre of \odwarf{} was substantially more s-process enriched than its outskirts. Some studies  \citep{Johnson2010,Smith2000,Marino2011,DaCosta2010} have analysed heavy-element abundances in \ocen{} and found strong star-to-star variations in s-process species together with evidence for a complex internal chemical-enrichment history. Their results are broadly consistent with what we find for the central region of \odwarf{}, where [La/Eu] decreases with radius and the most s-process-enhanced stars are concentrated toward the centre.

\begin{figure}[ht!]
    \centering
    \includegraphics[width=0.99\linewidth]{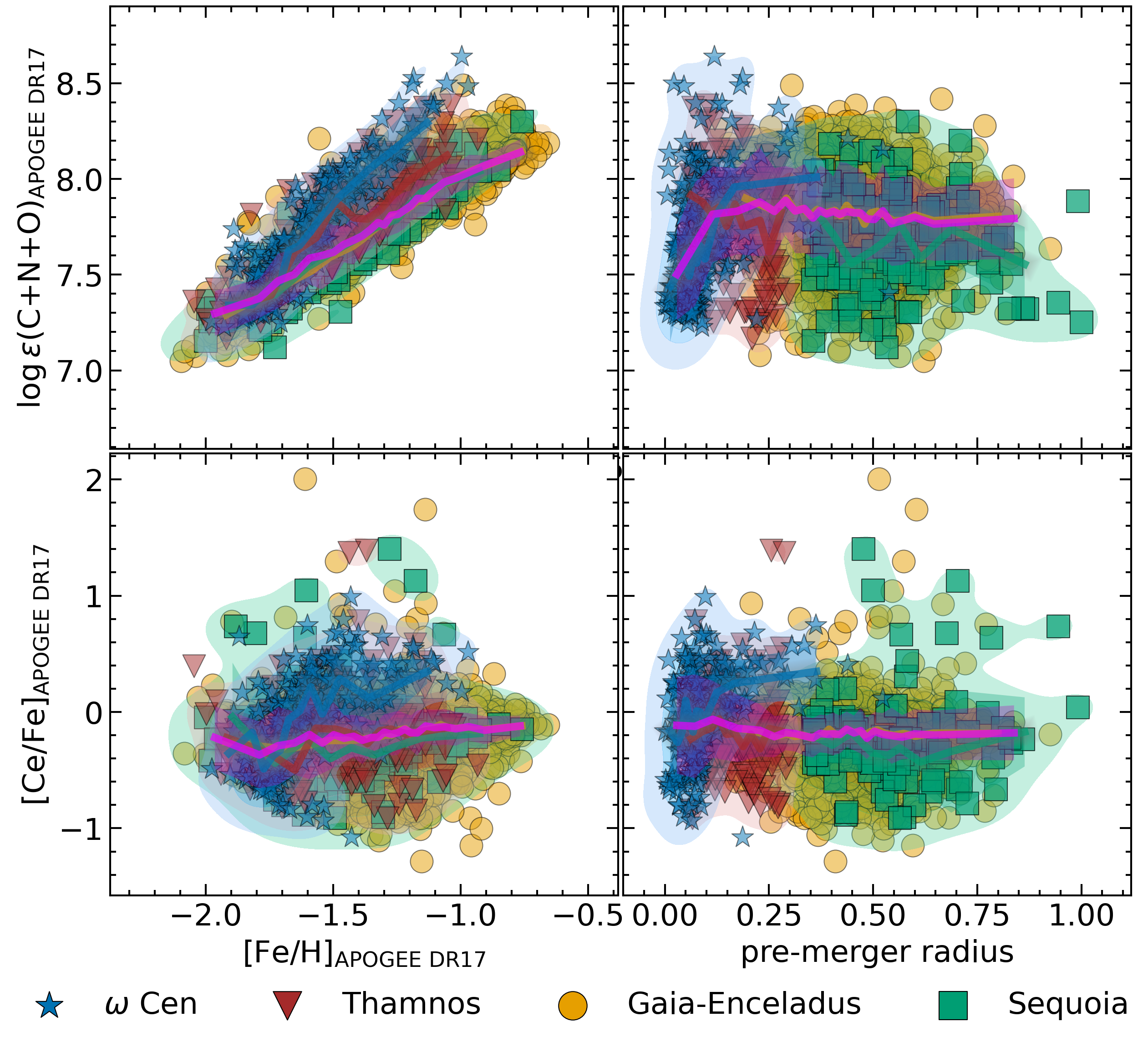}
    \caption{Total CNO and Ce abundance trends across \odwarf. Top panels show $\log\epsilon(\mathrm{C{+}N{+}O})$ versus [Fe/H] (left) and pre-merger radius (right), while the bottom panels show [Ce/Fe] versus [Fe/H] (left) and pre-merger radius (right). Symbols and colours are the same as in \autoref{fig:target_selection}, and the magenta line marks the median trend. The total CNO content mainly follows metallicity, whereas [Ce/Fe] shows only modest differences between components.}
    \label{fig:cno_ce_grad}
\end{figure}

In \autoref{fig:cno_ce_grad}, the absolute C$+$N$+$O abundance increases with [Fe/H] in all substructures, as also found by \cite{Marino2012} and \cite{Wang2026} for \ocen{}, suggesting that the buildup of the total CNO reservoir broadly followed the overall metallicity growth of the system. Likewise, [Ce/Fe], a heavy s-process element closely related to La, shows only modest differences among the \odwarf{} components. Taken together, these trends indicate that the clearest separation among the \odwarf{} components does not arise from C$+$N$+$O or [Ce/Fe] alone, but rather from the neutron-capture ratios [Ba/Eu] and [La/Eu]. While the metallicity and $\alpha$-element gradients are consistent with a formation pathway involving the merger of inspiraling GCs, the strong central enhancement in [Ba/Eu] and [La/Eu] requires substantial AGB feedback, implying extended star formation and a deep enough potential well to retain enriched gas in the inner regions (\ocen{}+Thamnos) \citep{Lanfranchi2005,Lanfranchi2008,Smith2000,Johnson2010, DOrazi2011}. This interpretation is consistent with the \ocen{} P2 formation scenario proposed by \cite{Dondoglio2025}. However, it remains unclear whether the same framework can also explain the enrichment pattern of the P1 population.

Because the main $s$-process contribution from low- and intermediate-mass AGB stars is delayed with respect to CCSN, and broadly comparable in timescale to the onset of SNe~Ia, the observed radial increase of [Ba/Eu] and [La/Eu] toward the inner \odwarf{} regions can be interpreted as a complementary tracer of the stripping sequence. In an outside-in disruption picture, the outer layers are removed first and therefore preserve a more chemically primitive composition, while the inner regions remain bound for longer, continue forming stars, and are more likely to retain slow AGB ejecta, thereby developing a stronger heavy-$s$ signature. In this sense, the enhanced $s$-process ratios in \ocen{} and, to a lesser extent, Thamnos are consistent with these components tracing material that survived longer in the progenitor potential than the more $r$-process-dominated outskirts associated with Sequoia and GE. Regardless of the statistic, the lower level of s-process in Sequoia compared to GE would support our proposed scenario where Sequoia is mainly stripped before GE. %Nevertheless, $s$-process enrichment is not a unique clock of tidal stripping, since it also depends on gas retention, mixing efficiency, metallicity-dependent AGB yields, and the local star-formation history.

\section{Summary}\label{sec:summary}

In this work, we tested the hypothesis that \ocen{}, Thamnos, Sequoia, and possibly Gaia--Enceladus are not independent accretion remnants, but instead trace different regions of a single disrupted progenitor dwarf galaxy, which we refer to as the \textit{Omega Dwarf} (\odwarf{}). To do so, we combined APOGEE DR17 and GALAH DR4 chemical abundances with \textit{Gaia} astrometry, constructed probabilistic memberships for the four components, and used present-day orbital energy as a proxy for pre-merger radius within the progenitor. Our main results are as follows:

\begin{itemize}
    \item The selected stars associated with \ocen{}, Thamnos, Sequoia, and GE occupy distinct but connected regions in orbital and chemical space, sharing some stars with the Nephele debris \citep{Pagnini2025b}. Their abundance distributions, even though offset, are broadly consistent in terms of relative distributions among the \odwarf{} components between APOGEE and GALAH, supporting the use of both surveys in a unified chemodynamical analysis.

    \item In the [Al/Fe]--[Fe/H] and [Mg/Mn]--[Al/Fe] planes, the P2 population, defined as He-N-Al-rich stars \citep[e.g.][]{Johnson2009, Johnson2010,Clontz2025,Wang2026}, is concentrated toward the inner regions, while the outskirts are dominated by the primordial P1 population -- a dwarf galaxy-like population. This suggests that the chemically evolved populations formed preferentially in the central part of the progenitor.

    \item When considering the primordial-like population (P1), the individual \odwarf{} components lie on the dwarf-galaxy sequence in the $\sigma(Z)^2$--$\bar{Z}$ plane \citep{Leaman2012}. This supports the interpretation that these structures are consistent with debris from a dwarf galaxy rather than with ordinary mono-metallic GCs.

    \item Using orbital energy as a proxy for pre-merger radius \citep{Carrillo2026}, we reconstruct a non-monotonic metallicity profile across \odwarf{}, with lower metallicity in the centre (\ocen{}+Thamnos), a maximum at intermediate pre-merger radius, and lower metallicity again in the outskirts, particularly in Sequoia, an inverted U-shaped gradient observed in similar mass nucleated dwarf galaxies \citep{Fahrion2021,Fahrion2022}.

    \item The [$\alpha$/Fe] gradient shows that the inner regions (\ocen{}+Thamnos) are more $\alpha$-enhanced than the outskirts (GE+Sequoia), indicating a more efficient early star formation in the central component typical of GCs. Taken together, the metallicity and $\alpha$-element gradients point out to a formation of the NSC P1 population via merger of inspiraling GCs \citep[][]{Guillard2016,Clontz2026, Wang2026}.

    \item The neutron-capture abundances reveal that the outer regions are more Eu-rich and show low [Ba/Eu] and [La/Eu], indicating a stronger relative r-process contribution. By contrast, the inner regions, especially \ocen{}, show enhanced [Ba/Eu] and [La/Eu], implying a stronger delayed s-process contribution from AGB stars \citep{Johnson2010, Dondoglio2025, Wang2026}. %The near-constant [Ba/La] ratio further indicates that the main variation is in the balance between s- and r-process enrichment, rather than within the heavy-s group itself.

    \item The absolute C$+$N$+$O abundance increases with metallicity in all components, while [Ce/Fe], [Ba/Eu], and [La/Eu] show strong s-process enhancement, which requires additional prolonged chemical evolution, efficient gas retention, and AGB feedback in the inner potential well. In this sense, a pure merger of inspiraling GC origin is insufficient to explain the full chemical complexity of the system, contrary to the metallicity and [$\alpha$/Fe] gradients.

    \item Even though Sequoia and Thamnos fit naturally into an outside-in stripping sequence around \ocen{}, the role of GE is less certain: although some of its chemical properties are compatible with an intermediate region of \odwarf{}, its action-space distribution remains harder to reconcile with the simplest version of the unified scenario. In this scenario, avoiding GE the mass of \odwarf{} tends to be lower than the one derived for GE \citep[$\sim 4\times10^9\, M_\odot$ recently derived by][]{Massari2026}.
\end{itemize}

Overall, the \odwarf{} scenario provides a coherent framework in which \ocen{} is the surviving nuclear star cluster of a disrupted dwarf galaxy, with GE being an intermediate region while Sequoia represents the first stars stripped during the merger and Thamnos a more recent disruption which potentially is connected with Fimbulthul stream \citep{Ibata2019}. This scenario can be tested by a multi-zone chemical evolution model assuming the energy as a proxy for pre-merger radius (Souza et al. in preparation).

The case for including GE remains plausible but not yet conclusive. Moreover, excluding GE from our proposed scenario does not change the chemical interpretation, only makes the dwarf galaxy slightly less massive, but still consistent with GE-like mass. To further test this hypothesis, more realistic N-body simulations with a surviving NSC are needed. More complete orbital information with the upcoming \textit{Gaia} DR4, and large spectroscopic datasets such as 4MOST \citep{deJong2019} will be essential to determine whether GE is truly part of the same progenitor or instead represents a separate accretion event partially overlapping with the \odwarf{} debris.

%% Please use the acknowledgment and contribution environments. This will 
%% be anonomyized when the "anonymous" style option is used. 
\begin{acknowledgements}

We thank Gail Zasowski for the insights, comments, and discussions during her stay at MPIA. Thank you also to Sebastian Kamann for the feedback and suggestions, as well as Sarah Martell and Rodrigo Ibata for the useful comments. We also thank Giulia Pagnini for having provided the Nephele APOGEE IDs. SOS acknowledges the support from Nadine Neumayer's Lise Meitner grant from the Max Planck Society and the DGAPA–PAPIIT grants IA103224 and IN112526. ACS, ZW, and CC acknowledge support from a \textit{Hubble Space Telescope} grant GO-16777. G.G. acknowledges support by Deutsche Forschungsgemeinschaft (DFG, German Research Foundation) – project-IDs: eBer-22-59652 (GU 2240/1-1"Galactic Archaeology with Convolutional Neural-Networks: Realising the potential of \textit{Gaia} and 4MOST"). AFK acknowledges funding from the Austrian Science Fund (FWF) [grant DOI 10.55776/ESP542]. This project has received funding from the European Research Council (ERC) under the European Union’s Horizon2020 research and innovation programme (Grant agreement No. 949173).
\end{acknowledgements}

%\begin{contribution}
% All authors contributed equally to the Terra Mater collaboration. (This is never true!)
% \end{contribution}
 
%\facilities{HST(STIS), Swift(XRT and UVOT), AAVSO, CTIO:1.3m, CTIO:1.5m, CXO}
 
\software{astropy \citep{Robitaille2013,PriceWhelan2018,TheAstropyCollaboration2022}, Numpy \citep{Harris2020}, Scipy \citep{Virtanen2020}, Matplotlib \citep{Hunter2007}, Seaborn \citep{Waskom2021}.}

\appendix
\counterwithin{figure}{section}
\counterwithin{table}{section}

\section{Dwarf galaxy-like population, using GALAH}

\begin{figure}
    \centering
    \includegraphics[width=0.995\linewidth]{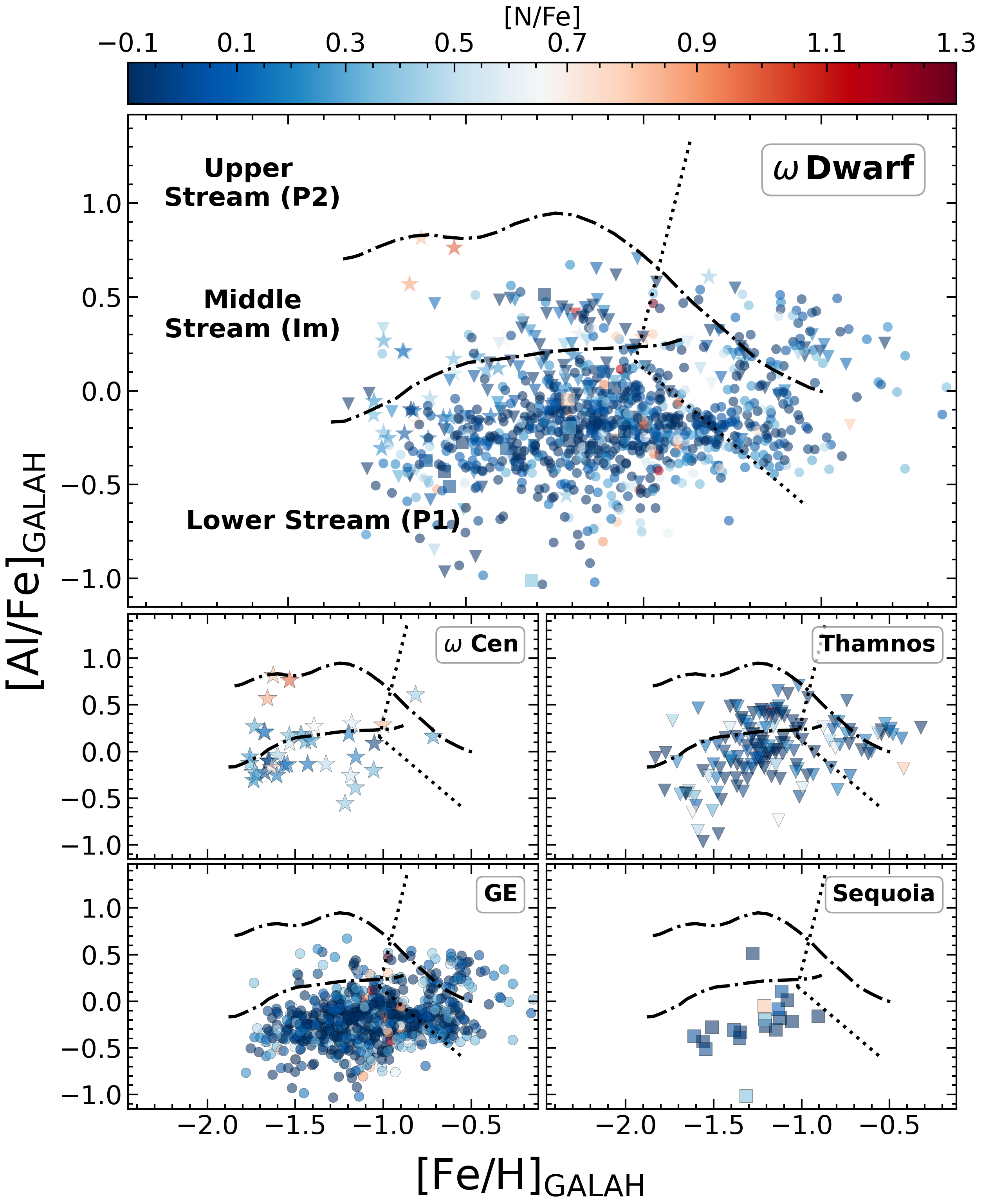}
    \caption{Populations across \odwarf{} using GALAH DR4 [Al/Fe] and [Fe/H] abundances. As in \autoref{fig:al_fe_apogee}, the upper panel represents the entire sample of \odwarf{} members, whose symbols and colours follow the same definition as in \autoref{fig:target_selection}. Each \odwarf{} substructure is individually displayed in the bottom panels. The lines show the limits for the lower (P1) and upper (P2) streams as defined by \cite{Dondoglio2025}, while the dotted lines show the limits for the unknown population.}
    \label{fig:al_fe_galah}
\end{figure}

\begin{figure*}[hbt!]
    \centering
    \includegraphics[width=0.99\linewidth]{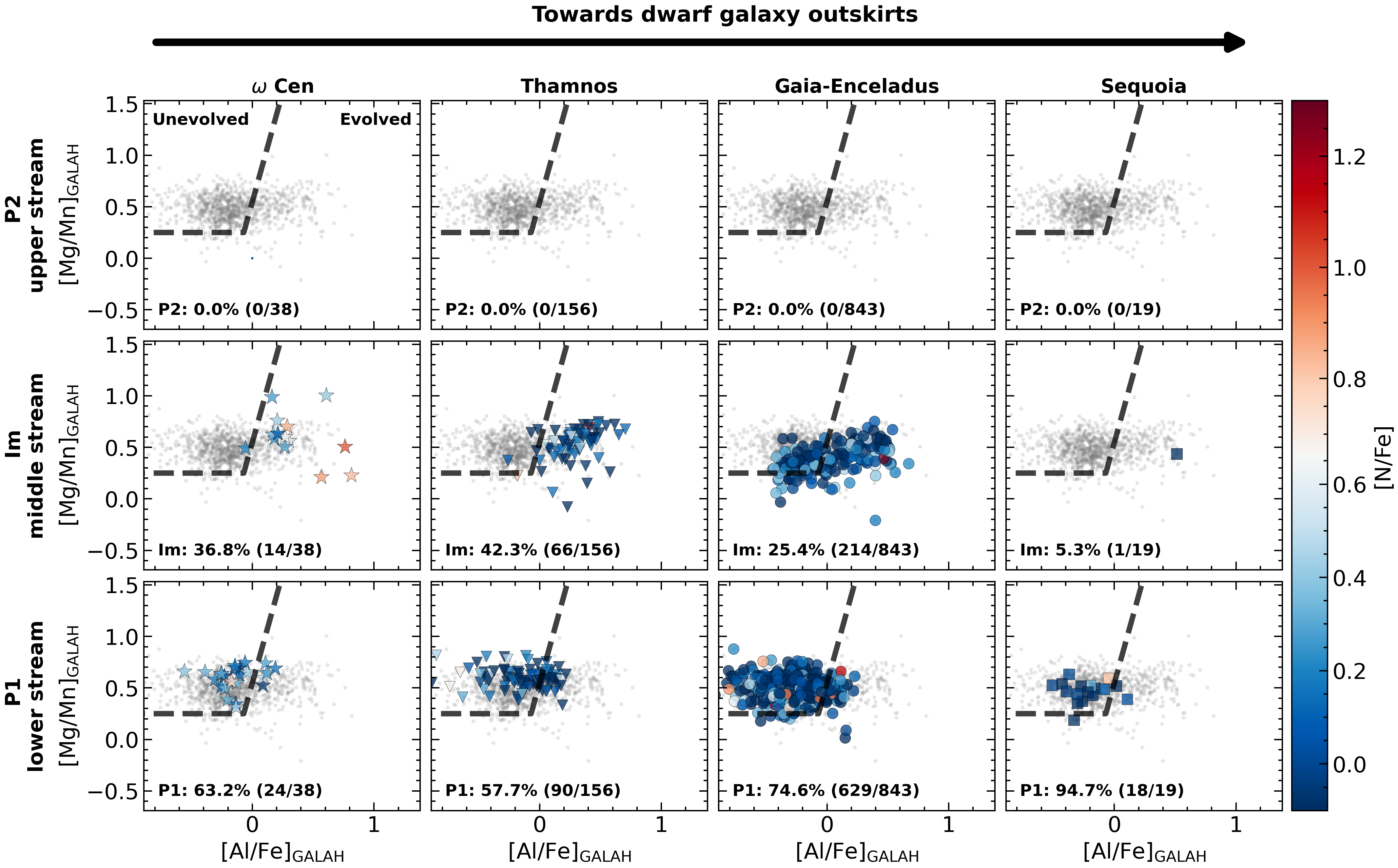}
    \caption{ \odwarf{} evolved and unevolved populations through the [Mg/Mn]--[Al/Fe] plane for GALAH. As in \autoref{fig:mgmn_apogee}, the components of \odwarf{} are individually displayed in the columns: \ocen{} in the left, Thamnos in the middle left, GE in the middle right, and Sequoia in the right. The top row represents the P2 population / upper stream, the middle stream in the central row, and the P1 population / lower stream in the bottom row. The color code indicates [N/Fe] values. The fraction (in percentage and absolute number) of each P2, Im, and P1 populations are in the bottom left.}
    \label{fig:mgmn_galah}
\end{figure*}

In \autoref{sec:dg_pop} we have discussed the populations across \odwarf{} using APOGEE abundances. Here we include the GALAH correspondent \autoref{fig:al_fe_galah} (\autoref{fig:al_fe_apogee}) and \autoref{fig:mgmn_galah} (\autoref{fig:mgmn_apogee}).

\bibliography{omegadwarf_cleaned}{}
\bibliographystyle{aasjournalv7}

\end{CJK}
\end{document}